\documentclass{PoS}
\usepackage{amsmath}
\input myunits.sty
\input mymath.sty
\usepackage{subfig}
\usepackage[sort&compress,square,comma,numbers,merge]{natbib}
\graphicspath{{./fig/}}
\DeclareGraphicsExtensions{.png,.pdf,.jpg,.mps}
\usepackage{grffile}

\title{Hybrid Mesons}

\ShortTitle{Hybrid Mesons}

\author{\speaker{Bernhard Ketzer}\\
        Physik Department, Technische Universit\"at M\"unchen,
        D-85748 Garching, Germany\\
        E-mail: \email{Bernhard.Ketzer@tum.de}}

\abstract{
  The SU(3)$_\mathrm{flavor}$ constituent quark model has been quite
  successful to explain the properties as well as the observed
  spectrum of mesons with pseudoscalar and vector quantum numbers. Many
  radial and orbital excitations of quark-antiquark ($q\overline{q}'$)
  systems 
  predicted by the model, however, have 
  not yet been observed experimentally or 
  assigned unambiguously.  
  In addition, a much richer spectrum of mesons is expected from QCD, in 
  which quarks interact which each other through the exchange of colored
  self-interacting gluons. Owing to this particular structure of QCD,
  configurations are allowed in which an excited gluonic field contributes to
  the quantum numbers $J^{PC}$ of the meson. States with a valence
  color-octet 
  $q\overline{q}'$ pair neutralized in color by an excited gluon field
  are termed hybrids. The observation of such states, however, is
  difficult because they will mix with ordinary $q\overline{q}'$  states
  with the same quantum numbers, merely augmenting the observed
  spectrum for a given $J^{PC}$. Since the gluonic field may carry
  quantum numbers other than $0^{++}$, however, this can give rise to
  states with ``exotic'' quantum numbers $J^{PC}=0^{--}$, $0^{+-}$, $1^{-+}$,
  $2^{+-},\ldots$ The lowest-lying hybrid multiplet is expected to
  contain a state with exotic
  quantum numbers $J^{PC}=1^{-+}$. 
  The identification of such a state is considered a ``smoking gun''
  for the observation of non-$q\overline{q}'$ mesons. 
  The
  search for hybrid states 
  has been a central goal of hadron spectroscopy in the last 20
  years. Ongoing and upcoming high-statistics
  experiments are expected to 
  shed new light on the existence of such states in nature. 
  In this paper, theoretical predictions for masses and decay modes as
  well as recent  
  experimental evidence for hybrid meson states 
  and future experimental 
  directions are discussed.  
}

\FullConference{Sixth International Conference on Quarks and Nuclear Physics,\\
		April 16-20, 2012\\
		Ecole Polytechnique, Palaiseau, Paris}

\begin{document}

\section{Introduction}
\label{sec:intro}
Mesons are experimentally characterized by quantum numbers
$I(J^{P})$,  
with the isospin $I$, 
the total
angular 
momentum $J=0,1,2\ldots$, and the parity $P$. Neutral flavorless
mesons are 
eigenstates of the particle-antiparticle conjugation operator and thus
have a defined 
charge conjugation parity $C$. Flavorless mesons are 
eigenstates of the $G$ parity. In the 
%SU(3)$_\mathrm{flavor}$ 
constituent quark model, 
mesons are described as color-singlet bound states of a quark $q$ and
an antiquark 
$\overline{q}'$. 
The quantum numbers $P$, $C$ and $G$ are then given by 
\begin{equation}
  \label{eq:meson.qn}
  P=(-1)^{L+1},\quad C=(-1)^{L+S},\quad G=(-1)^{L+S+I},
\end{equation}
where $L$ is the relative orbital angular momentum of $q$ and
$\overline{q}'$, 
and $S$ the total intrinsic spin of the $q\overline{q}'$ pair, with
$S=0,1$. 
This gives rise to meson states with quantum numbers $J^{PC}=0^{-+}$, 
$0^{++}$, $1^{--}$, $1^{+-}$, $1^{++}$, $2^{--}$, $2^{-+}$, $2^{++},\ldots$. 

Quantum Chromodynamics (QCD) allows for a much richer spectrum of
mesons, including  
configurations in which an excited gluonic field
contributes to 
the quantum numbers $J^{PC}$ of the meson. States with a valence
color-octet 
$q\overline{q}'$ pair neutralized in color by an excited gluon field
are termed hybrids. Since the gluonic field may carry
quantum numbers different from those of the vacuum, hybrids
can also appear 
with ``exotic'' quantum numbers $J^{PC}=0^{--}, 0^{+-}, 1^{-+},
2^{+-},\ldots$. 
%The lowest-lying hybrid multiplet is expected to
%contain a state with exotic
%quantum numbers $J^{PC}=1^{-+}$ \cite{Dudek:2011bn}. 
The unambiguous identification of resonances with such quantum
numbers is considered a ``smoking gun'' for the observation of mesons
beyond the $q\overline{q}'$ configuration. Recent
and upcoming high-statistics 
experiments are about or expected to 
shed new light on the existence of such states in nature. 
In this paper 
theoretical predictions and the present experimental evidence for
hybrid meson states are discussed.
%\cite{Nozar:2008bea,Alekseev:2009xt}
Directions for and expectations from future experiments 
are given.  

\section{Theoretical Predictions}
\label{sec:theory}
%\subsection{QCD}
The properties of hadrons 
are traditionally estimated from phenomenological models like the
quark model, 
the bag model, 
the flux tube model, QCD sum rules, or effective theories with pions
and other 
pseudoscalar 
particles as the 
active degrees of freedom. Recently, also lattice QCD
(LQCD) started to make 
predictions for meson  
properties 
like their masses and widths, which can be tested by
experiments.    

\subsection{Models}
\label{sec:theory.models}
%\subsubsection{Bag Model}
In the Bag model %\cite{Chodos:1974je,DeGrand:1975cf},  
gluonic excitations
correspond to modes of the gluonic field allowed by
the boundary conditions on the surface of the bag. The lowest
eigenmodes are    
transverse electric (TE) and transverse magnetic (TM) gluons  
with quantum
numbers $J^{PC}=1^{+-}$ and $1^{--}$, respectively
\cite{Jaffe:1975fd}. 
The lowest-mass hybrid mesons are formed by
combining a 
$q\overline{q}'$ system with 
spin $0$ or $1$ with a TE gluon. This yields four nonets with quantum
numbers and masses given in Table~\ref{tab:hybrid_qn}.
\begin{table}[tbp]
  \caption{Quantum numbers and approximate masses of non-strange
    hybrids 
    in various models. The LQCD masses in parentheses are given for a 
    pion mass of $396\,\MeV/c^2$. $J^{PC}_{q\overline{q}'}$
    and $J^{PC}_g$ denote 
    the quantum numbers of the $q\overline{q}'$ pair and the gluon,
    respectively, which couple to give $J^{PC}$ of the hybrid. Bold
    numbers indicate spin-exotic quantum numbers.}
  \label{tab:hybrid_qn}
  \centering
  {\small 
  \begin{tabular}{ccccc} \hline\hline
    Model & $J^{PC}_{q\overline{q}'}$ & $J^{PC}_{g}$ & $J^{PC}$ & Mass
    ($\GeV/c^2$) \\ \hline
    Bag \cite{Chanowitz:1982qj,Barnes:1982tx} 
    & $0^{-+}$ & $1^{+-}$ (TE) & $1^{--}$ & $\sim 1.7$ \\
        & $1^{--}$ & $1^{+-}$ (TE) & $(0,{\mathbf 1},2)^{-+}$ & 
        $\sim 1.3,1.5,1.9$\\ 
% wrong TM quantum numbers! (from Meyer et al.)        
%        & $0^{-+}$ & $1^{-+}$ (TM) & $1^{++}$ & heavier \\
%        & $1^{--}$ & $1^{-+}$ (TM) & $({\mathbf 0}, 1, {\mathbf
%          2})^{+-}$ & heavier \\ \hline    
% correct TM quantum numbers (from Chanowitz et al.)
        & $0^{-+}$ & $1^{--}$ (TM) & $1^{+-}$ & heavier \\
        & $1^{--}$ & $1^{--}$ (TM) & $(0,1,2)^{++}$ & heavier \\ \hline    
    Flux tube \cite{Isgur:1984bm,Barnes:1995hc} 
        & $0^{-+}$ & $1^{+-}$ & $1^{--}$ & $1.7$-$1.9$ \\
        & $1^{--}$ & $1^{+-}$ & $(0,{\mathbf 1},2)^{-+}$ & $1.7$-$1.9$ \\
        & $0^{-+}$ & $1^{-+}$ & $1^{++}$ & $1.7$-$1.9$  \\
        & $1^{--}$ & $1^{-+}$ & $({\mathbf 0}, 1, {\mathbf
          2})^{+-}$ & $1.7$-$1.9$  \\ \hline    
    Constituent gluon & $0^{-+}$ & $1^{--}$ & $1^{+-}$ & $1.3$-$1.8$ /
    $2.1$ \\  
    \cite{Ishida:1991mx}/\cite{General:2006ed}
     & $1^{--}$ & $1^{--}$ & $(0,1,2)^{++}$ & $1.3$-$1.8$ / $2.2$ \\  
         & $1^{+-}$ & $1^{--}$ & $(0,{\mathbf 1},2)^{-+}$ &
         $1.8$-$2.2$ / $2.2$\\  
         & $(0,1,2)^{++}$ & $1^{--}$ &
         $1^{--},({\mathbf 0},1,2)^{--},(1,2,3)^{--}$ & $1.8$-$2.2$ / $2.3$ \\ \hline
    Constituent gluon / 
         & $0^{-+}$ & $1^{+-}$ & $1^{--}$  
         & $(2.3)$ \\                %% lin.extrapol. $(2.3)$ \\  
    LQCD \cite{Guo:2007sm,Dudek:2011bn} 
         & $1^{--}$ & $1^{+-}$ & $(0,{\mathbf 1},2)^{-+}$ 
         & $(2.1,2.0,2.4)$ \\%% li. extrapol. $(2.0,1.8,2.3)$ \\   
         & $1^{+-}$ & $1^{+-}$ & $(0,1,2)^{++}$ & $(>2.4)$ \\  
         & $(0,1,2)^{++}$ & $1^{+-}$ &
         $1^{+-},({\mathbf 0},1,{\mathbf 2})^{+-},(1,{\mathbf
           2},3)^{+-}$ & $(>2.4)$ \\
%     LQCD & & & $1^{--}$ & $2.1$ - $2.3$ \\  
%          & & & $(0,{\mathbf 1},2)^{-+}$ & $2.1$ - $2.3$ \\  
%          & & & $(0,1,2)^{++}$ & \\  
%          & & & $1^{+-},({\mathbf 0},1,{\mathbf 2})^{+-},(1,{\mathbf
%            2},3)^{+-}$ & \\ \hline
        \hline\hline    
  \end{tabular}
}
\end{table}

%\subsubsection{QCD Spectral Sum Rules}
Calculations using the QCD 
spectral
sum rule (QSSR) approach favor the $1^{-+}$ state at a mass 
between\ $\lesssim 1.5\,\GeV/c^2$ \cite{Balitsky:1986hf}, and $\sim
1.8\,\GeV/c^2$ 
% with dominant decay widths to $\rho\pi$
%($274\,\MeV/c^2$), $\gamma\pi$ ($3\,\MeV/c^2$), $\eta'\pi$ ($3\,\MeV/c^2$)
\cite{Latorre:1985tg,Chetyrkin:2000tj,Narison:2009vj}. 
The latter 
authors also find a $1^{--}$ hybrid almost degenerate with the
$1^{-+}$, while they expect the exotic $0^{--}$ at a much higher mass
of $\sim 2.8\,\GeV/c^2$. Other low-mass hybrids have not been
investigated in more detail in the QSSR approach.  
%found an exotic 
%$1^{-+}$ state at low masses around $1\,\GeV/c^2$, while later
%calculations yielded higher masses between $1.6$ and $2.1\,\GeV/c^2$
%\cite{Latorre:1984kc,Latorre:1985tg}.   

%\subsubsection{Flux Tube Model}
In the flux tube model %\cite{Isgur:1984bm,Isgur:1985vy} 
gluonic excitations 
correspond to transverse vibrations of the string-like flux tube
between a $q\overline{q}'$ pair. 
For zero angular momentum, the flux tube behaves as if it has quantum
numbers $J_g^{PC}=0^{++}$, which results in the ordinary quark model
meson quantum numbers when combined with the underlying
$q\overline{q}'$ system. For one unit of angular momentum, the flux
tube quantum numbers can be $J_g^{PC}=1^{-+}$ or $1^{+-}$. 
The quantum
numbers and masses of the gluonic 
excitations as predicted by the flux tube model are summarized in
Table~\ref{tab:hybrid_qn}. 

%\subsubsection{Constituent Gluons}
An alternative description of gluonic excitations suggests to
construct Fock states of hadrons from massive constituent quarks and
gluons. 
In the simplest constituent gluon picture, %\cite{Horn:1977rq},  
a transverse quasigluon 
with $J^{PC}=1^{--}$ is added in a relative $S$ wave to a
$q\overline{q}'$ system in an $S$ wave with spin $0$ or $1$. This
gives 
rise to a 
supermultiplet of four nonets with the same quantum numbers as $P$ wave
$q\overline{q}'$ states, none of them exotic, as 
shown in Table~\ref{tab:hybrid_qn}. 
Predictions for the masses of the non-strange members of these nonets
vary considerably between different approaches withing the constituent
gluon picture. 

\subsection{Lattice Gauge Theory}
\label{sec:theory.lqcd}
\begin{figure}[tbp]
  \begin{center}
    \includegraphics[width=0.9\columnwidth]{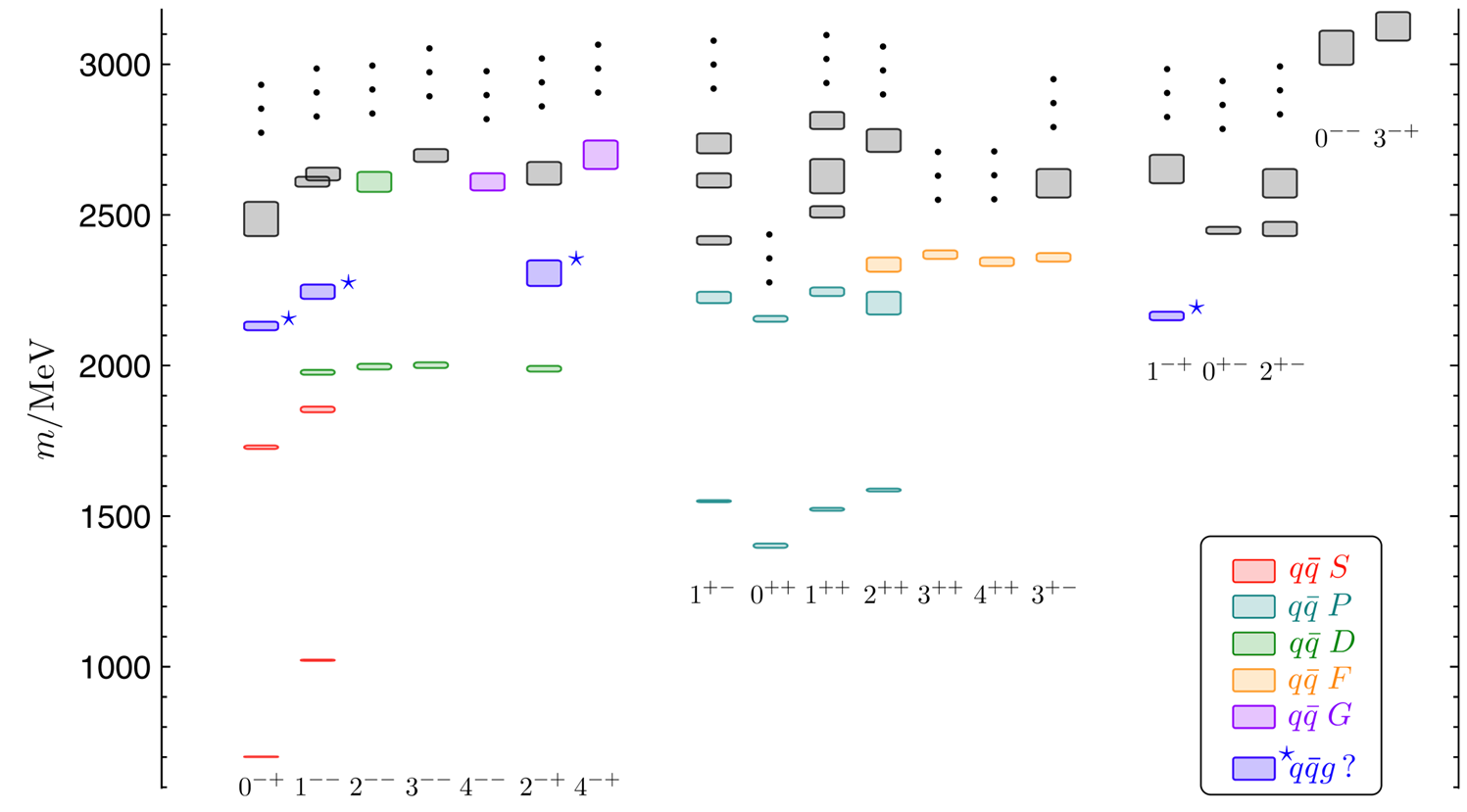}
    \caption{Spectrum of light isovector mesons resulting from a
      state-of-the-art lattice calculation \cite{Dudek:2011bn}, sorted
      by their quantum numbers $J^{PC}$. The box size indicates the
      statistical uncertainty on the masses. The
      colors indicate the dominant structure of the states. The pion
      mass used in these calculations is $700\,\MeV/c^2$.}
    \label{fig:lqcd_isovector_spectrum}
  \end{center}
\end{figure}

Most LQCD studies focused on the exotic $1^{-+}$ hybrid. 
%, and give a mass of
%$\sim 1.9\,\GeV/c^2$ \cite{Michael:2003xg}. 
A more recent quenched 
calculation on large lattices, performed with 
pion masses as light as $320\,\MeV/c^2$ gives a mass of the $1^{-+}$
exotic of 
$1.74(25)\,\GeV/c^2$ when extrapolated to the physical pion mass 
\cite{Hedditch:2005zf}. 
Recently, a fully dynamical (unquenched) calculation of
the complete spectrum of light-quark isoscalar and isovector mesons
has been performed. %\cite{Dudek:2010wm,Dudek:2011tt}.  
The resulting isovector
meson spectrum for a pion mass of $700\,\MeV/c^2$ is depicted in
Fig.~\ref{fig:lqcd_isovector_spectrum} \cite{Dudek:2011bn}. 
Quantum numbers and the quark-gluon structure of a meson state
$n$ with a given
mass $m_{n}$ are extracted by studying matrix elements 
$\langle n|\mathcal{O}_i|0\rangle$, which encode the extent
to which operator $\mathcal{O}_i$ overlaps with state $n$.  
The dominant operator overlaps are indicated by
the colors, and the corresponding operators are indicated in
suggestive form in the lower right hand corner of
Fig.~\ref{fig:lqcd_isovector_spectrum}.  

In addition to ordinary states
consisting of a valence $q\overline{q}$ pair, the calculation also
predicts states 
in which both quark and gluonic degrees of freedom contribute to the
quantum numbers (hybrids). The lowest-lying isovector hybrid
states form a 
super-multiplet consisting of a spin-exotic
state with $J^{PC}=1^{-+}$, and three non-exotic states with
quantum numbers $0^{-+}$, $1^{--}$, and $2^{-+}$. 
%The first excited
%hybrid states according to this calculation are orbital excitations of
%the $q\overline{q}'$ pair with the same gluonic excitation as the
%ground states. 

It is interesting to
note that none of the models described in Sec.~\ref{sec:theory.models}
is able to reproduce the 
degeneracy pattern of hybrid states emerging from LQCD. Considering a
model, in which a constituent gluon couples to a 
$q\overline{q}'$ pair in a $P$ wave rather than in an $S$ wave as
above, such that $J^{PC}=1^{+-}$, one is able to successfully
reproduce the pattern of hybrid  
states from LQCD \cite{Dudek:2011bn,Guo:2007sm}, as shown in
Table~\ref{tab:hybrid_qn}. 

\subsection{Decay Modes}
\label{sec:theory.decay}
In order to clarify the nature of a non-$q\overline{q}'$ resonance, it is
important to study its decay properties and compare them to
theoretical predictions. Hybrid meson decays have been studied in
models \cite{Isgur:1985vy,Close:1994hc,Page:1998gz} as
well as, recently, in LQCD \cite{McNeile:2006bz}. The 
results are 
are summarized in 
Table~\ref{tab:hybrid.decay}
for low-mass hybrids with
$J^{PC}=1^{-+},0^{-+},1^{--},2^{-+}$.  
\begin{table}[tbp]
  \caption{Total widths in $\MeV/c^2$ and dominant decay
    channels of light isovector hybrid mesons
    according to model 
    calculations and LQCD.
    Values labeled IKP 
    \cite{Isgur:1985vy} and PSS \cite{Page:1998gz} are for a hybrid mass 
    of $1.8\,\GeV/c^2$. Results for the IKP model have been
    calculated in \cite{Page:1998gz}. LQCD \cite{McNeile:2006bz}
    results are for a 
    hybrid mass of $2.0\,\GeV/c^2$. 
  }  
  \label{tab:hybrid.decay}
  \centering{\small
  \begin{tabular}{ccccc} \hline\hline
    $J^{PC}$ & \multicolumn{3}{c}{Total width ($\,\MeV/c^2$)} 
    & Dominant channels \\ 
    & IKP \cite{Isgur:1985vy} & PSS
    \cite{Page:1998gz} & LQCD \cite{McNeile:2006bz} & \\ \hline 
    $1^{-+}$ & 117 & 81 & 180 
    & $b_1(1235)\pi$, $f_1(1285)\pi$, $\rho\pi$, $\eta(1295)\pi$,
    $a_1(1260)\eta$ \\
    $0^{-+}$ & 132 & 102 & 
    & $f_0(1370)\pi$, $\rho\pi$, $\rho(1450)\pi$, $K^* K$ \\
    $1^{--}$ & 112 & 70 & 
    & $a_1(1260)\pi$, $a_2(1320)\pi$, $\omega\pi$, $K_1(1270)K$,
    $\rho\eta$ \\ 
    $2^{-+}$ & 59 & 27 & 
    & $f_2(1270)\pi$, $b_1(1235)\pi$, $\rho\pi$ \\
    \hline\hline
  \end{tabular}
}
\end{table}

% According to flux tube model calculations \cite{Ackleh:1996yt}, light
% mesons 
% preferably 
% decay via the creation of a $q\overline{q}'$ pair with vacuum quantum
% numbers $J^{PC}=0^{++}$, i.e.\ $L=1$, $S=1$ ($^3P_0$ model
% \cite{Micu:1968mk}). For hybrid mesons, most models predict that
% decays into an $L=0$ and an $L=1$ meson are  
% preferred over decays to two $L=0$ states \cite{Close:1994hc}. The
% selection rule 
% is only approximate, however, and broken if the newly created mesons
% have different spatial wavefunctions, as is the case for the decay
% into $\rho\pi$.   
% According to SU(3)$_{\mathrm{flavor}}$ symmetry considerations
% \cite{Chung:2002fz}, a $1^{-+}$ hybrid meson should prefer decay via 
% $\eta'\pi$ rather than $\eta\pi$. 

\section{Experiments}
\label{sec:exp}
Experimental evidence for the existence of hybrid mesons can come from
two sources. The observation of an overpopulation of states with
$q\overline{q}'$ quantum numbers may indicate the existence of 
states beyond the quark model, i.e.\ hybrids, glueballs or multi-quark
states. The densely populated spectrum of light mesons in the mass
region between $1$ and $2\,\GeV/c^2$, and the broad nature of the
states involved, however, makes this approach difficult. 
The identification of a resonance with exotic quantum numbers,
however, is a clear evidence 
for the existence of such states. 
Unfortunately, models and LQCD predictions do not agree on 
the quantum numbers and masses expected for the
lowest-mass hybrids (c.f. Sec.~\ref{sec:theory}).  
It is
thus of great importance for our understanding 
of non-perturbative QCD to test those
predictions experimentally.   

\subsection{Production Mechanisms}
\label{sec:exp.production}
Production experiments, where the total energy is shared
between a multi-meson final state $X$ and a recoil particle, can be
employed to produce both non-exotic and exotic states. The quantum
numbers of the multi-meson system are restricted 
only by conservation laws of the reaction. The final state
will contain contributions from all (indistinguishable) intermediate
states with different quantum numbers. A partial wave analysis is
usually required to disentangle the different contributions. 
The decay of a resonance into a multi-particle final state is 
commonly described in the isobar 
model, which assumes a series of sequential two-body decays into
intermediate resonances 
(isobars), which eventually decay into the final state observed in the
experiment.  
A partial wave is characterized by a set of
quantum numbers  
$J^{PC}M^\epsilon\,R_1R_2\,\left[L\atop S\right]$; 
$M$ is the absolute value of the spin projection onto the
$z$-axis; $\epsilon$ is the
reflectivity;
% \cite{Chung:1974fq}, which describes the symmetry under a
% reflection through the 
% production plane; 
%, and which is 
%defined such that it 
%corresponds to the naturality of
%the exchanged Regge trajectory; 
$L$ is the orbital angular momentum
between the isobars $R_1$ and $R_2$, and $S$ their total spin. 

\begin{figure}[tbp]
  \begin{center}
    \hfill
    \subfloat[]{
      \includegraphics[width=0.2\columnwidth]{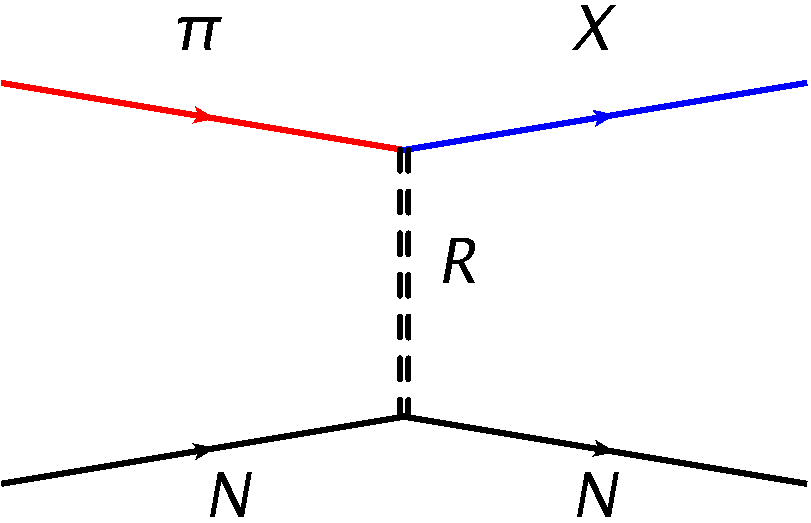}
      \label{fig:exp.production.diff}}
    \hfill
    \subfloat[]{
      \includegraphics[width=0.3\columnwidth]{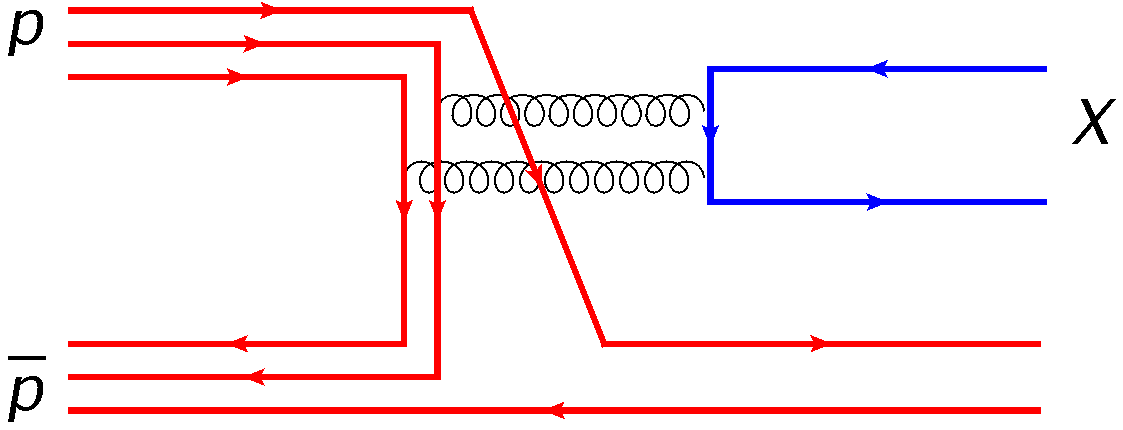}
      \label{fig:exp.production.pbarp}}
    \hfill
    \subfloat[]{
      \includegraphics[width=0.2\columnwidth]{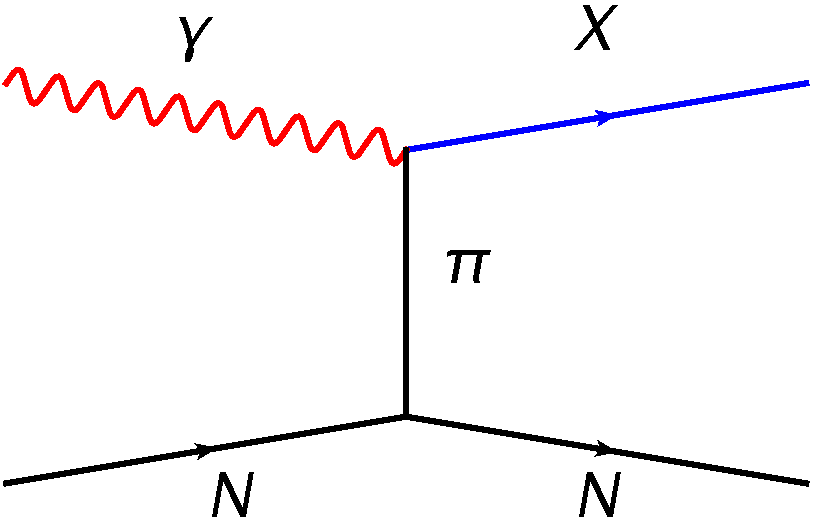}
      \label{fig:exp.production.photo}}
    \hfill
    \subfloat[]{
      \includegraphics[width=0.2\columnwidth]{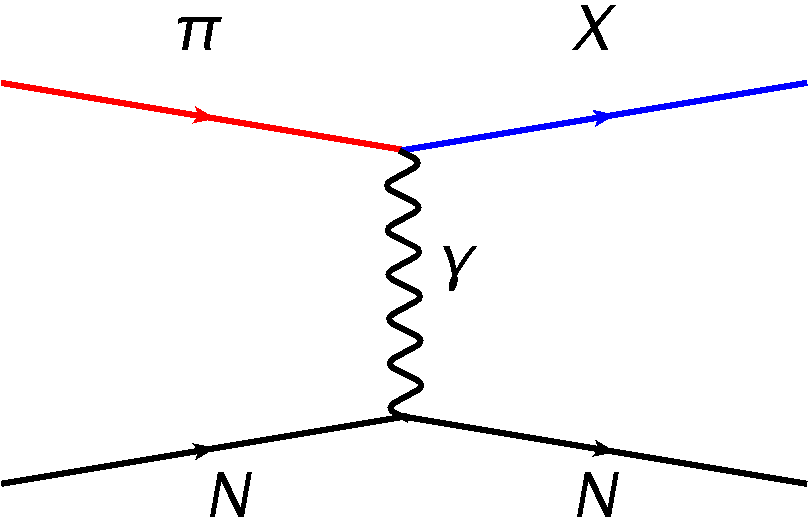}
      \label{fig:exp.production.coulomb}}
    \hfill
      \caption{Production mechanisms of hybrid mesons.}
      \label{fig:exp.production}
    \end{center}
  \end{figure}

The most
prominent production reactions are shown in
Fig.~\ref{fig:exp.production}. Diffractive excitation
(\ref{fig:exp.production.diff}) of an incoming  
beam particle (usually a pion or a kaon) proceeds via the exchange of
a Regge 
trajectory $R$ in 
the $t$-channel. Angular momentum as well as 4-momentum $t$ is
exchanged. Instead of $t$, the positive variable 
$t' = \vert t \vert - \vert t \vert_\mathrm{min}$, 
% = 2\vert\vec{p}_a\vert \vert\vec{p}_c\vert (1-\cos\theta_0) \geq 0$, 
is often used, 
where
$\vert t \vert_\mathrm{min}$ is 
the minimum value of $\vert t \vert$ which 
is needed by kinematics to produce a final state $X$ of a given
mass.  
Antiproton annihilation (\ref{fig:exp.production.pbarp}) can
proceed either in flight or at 
rest, and on protons or on neutrons. For $\overline{p}p$, the initial
state is a mixture of isospin $I=0$ and $1$, with $I_3=0$, while for
$\overline{p}n$ the initial state is pure $I=1$. 
Photoproduction proceeds via real or quasi-real virtual
photons. Using a beam of real photons (\ref{fig:exp.production.photo})
with quantum numbers 
$J^{PC}=1^{--}$, the production of spin-exotic hybrids with $S=1$ is
expected to  
be favored from Vector Meson Dominance, compared to diffractive
production from a pion beam with $J^{PC}=0^{-+}$ and $S=0$. 
%According to models, a spin-exotic resonance could be produced with a
%strength comparable to the $a_2(1320)$. 
This
hypothesis can also be tested using quasi-real photons
(\ref{fig:exp.production.coulomb}) from the
Coulomb field of a heavy target nucleus. 
In addition, hard exclusive leptoproduction was also proposed
as a tool to study a $1^{-+}$ hybrid \cite{Anikin:2004ja}. 

\subsection{Hybrids with Exotic Quantum Numbers}
\label{sec:exp.exotic}
Models as well as LQCD consistently predict the lightest spin-exotic
hybrid to have quantum numbers $J^{PC}=1^{-+}$. 
Currently, there are 
three 
experimental candidates for a light $1^{-+}$ hybrid 
\cite{Beringer:2012zz}: the
$\pi_1(1400)$ and the $\pi_1(1600)$, observed in diffractive reactions
and $\overline{p}N$ annihilation, and the $\pi_1(2015)$, seen only in
diffraction. 
The resonant nature of these states is 
still heavily disputed in the 
community (for a recent review, see \cite{Meyer:2010ku}). 
New experiments with higher statistics and better acceptance, allowing
for more elaborate analysis techniques, are
needed in order to shed new light on these questions. 

\subsubsection{Diffractive Production}
\label{sec:exp.exotic.diffraction}
The COMPASS experiment \cite{Abbon:2007pq} at CERN's Super Proton
Synchrotron (SPS) is a 
large-accep\-tance, high-resolution magnetic spectrometer, which is
gathering high-statistics event samples of diffractive and Coulomb
production reactions of hadronic beam particles into
final states 
containing charged and neutral particles. 

\begin{figure*}[tbp]
  \begin{center}
    \subfloat[]{
      \includegraphics[width=0.45\columnwidth]{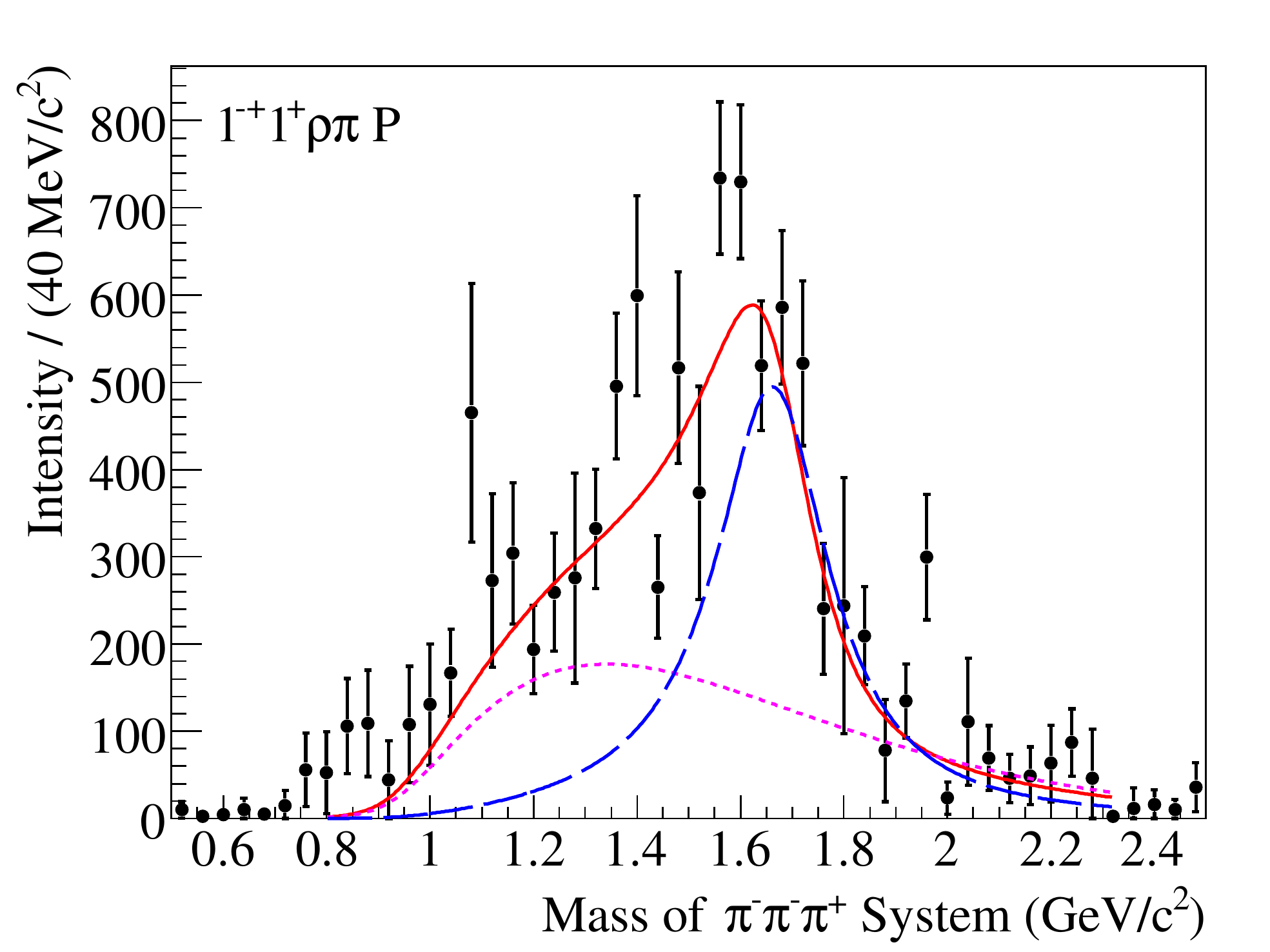}
      \label{fig:Pb.3pi.1-+.intensity}}
    % \subfloat[]
    %   \includegraphics[width=0.3\columnwidth]{1-+_1+_rho_pi_P_2008lH}}
    \subfloat[]{
      \includegraphics[width=0.45\columnwidth,height=0.35\columnwidth]{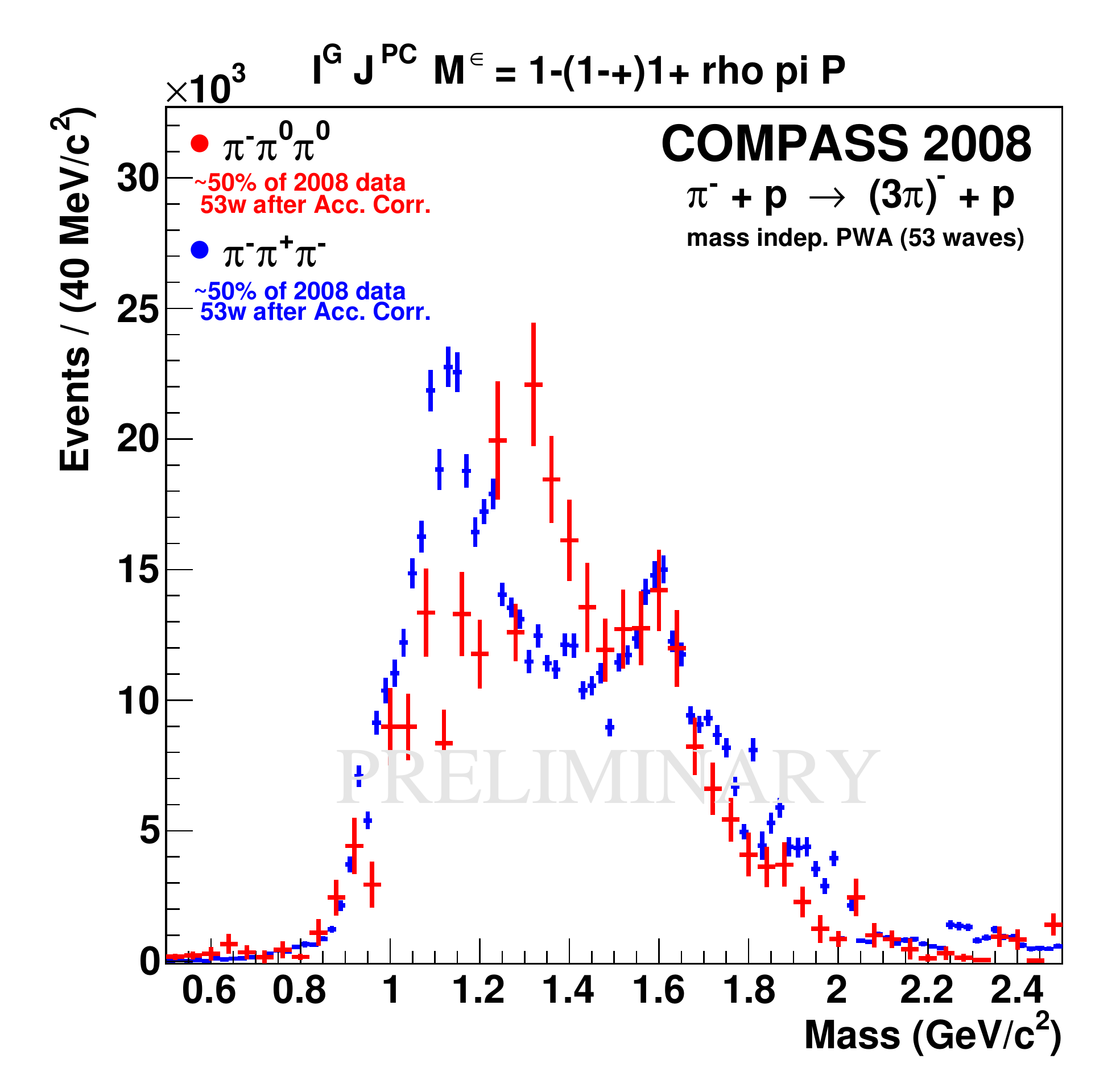}
      \label{fig:H.3pi.1-+.intensity}}
    \caption{Intensity of exotic $1^{-+}1^+\,\rho\pi\,P$ wave as a
      function of $3\pi$ invariant mass for 4-momentum transfer between
      $0.1$ and $1.0\,\GeV^2/c^2$, (a) for Pb
      target and $\pi^-\pi^-\pi^+$ final state, (b) for H target 
      and charged (blue) and
      neutral (red) $3\pi$ final states.} 
    \label{fig:3pi.1-+.intensity}
  \end{center}
\end{figure*}

\begin{figure*}[tbp]
  \begin{center}
    \subfloat[]{
      \includegraphics[width=0.45\columnwidth]{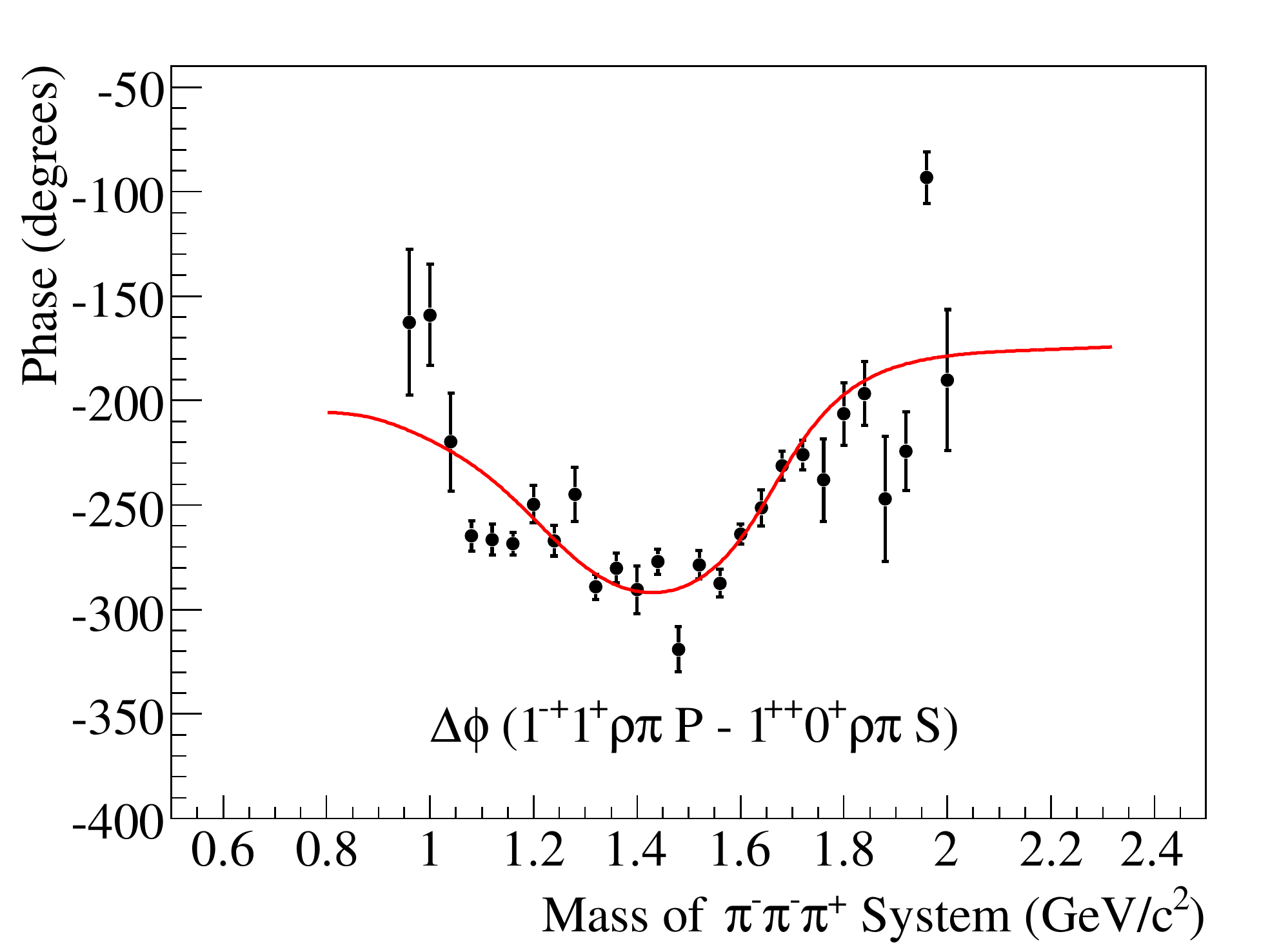}
      \label{fig:Pb.3pi.1-+.phase}}
%    \subfloat[]{
      % \includegraphics[width=0.3\columnwidth]{phase_1-+_1+_rho_pi_P_vs_1++_0+_rho_pi_S_2008lH}}
    \subfloat[]{
      \includegraphics[width=0.45\columnwidth, height=0.35\columnwidth]{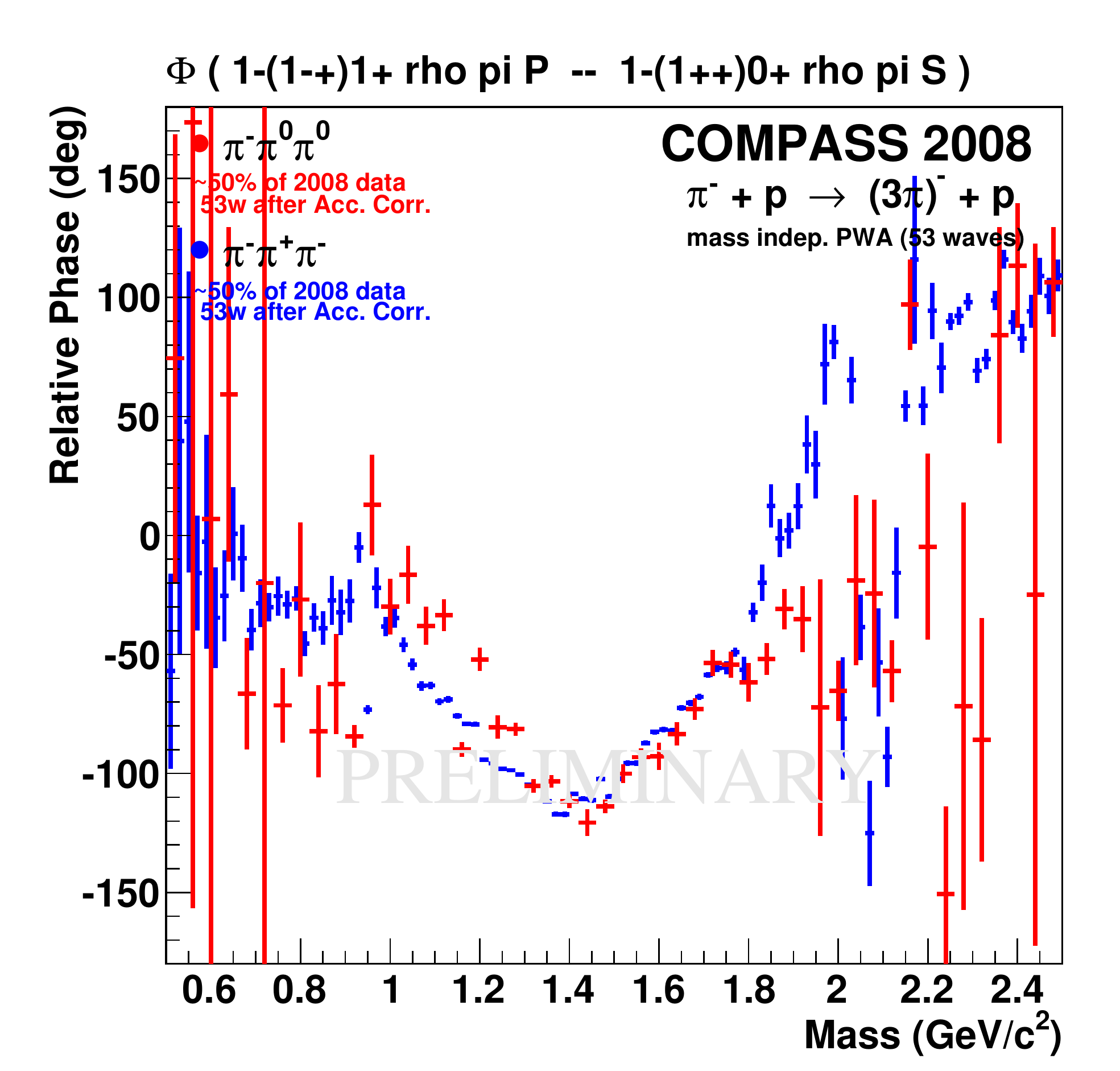}
      \label{fig:H.3pi.1-+.phase}}
      \caption{Phase difference between the $1^{-+}1^+\,\rho\pi\,P$
        and the $1^{++}0^+\,\rho\pi\,S$ waves, (a) for Pb target and 
        $\pi^-\pi^-\pi^+$ final state, (b) for H target and charged (blue) and
        neutral (red) $3\pi$ final states.} 
    \label{fig:3pi.1-+.phase}
  \end{center}
\end{figure*}

\paragraph{$3\pi$ Final State}\mbox{}\\
\indent In a first analysis of
$420\,\mathrm{k\ events}$ of the $\pi^-\pi^-\pi^+$ final state from
scattering of 
$190\,\GeV/c$ $\pi^-$
on a Pb target with 
4-momentum transfer $0.1<t'<1.0\,\GeV^2/c^2$, a clear signal in the
$1^{-+}1^+\,\rho\pi\,P$ partial wave has been
observed \cite{Alekseev:2009xt}, as shown in
Fig.~\ref{fig:Pb.3pi.1-+.intensity}.  
The data points are the results of independent extended maximum
likelihood fits of production amplitudes to angular
distributions in bins of the
$3\pi$ mass, including 41 partial waves and a flat background wave. 
The curves show a fit of
Breit-Wigner resonances (blue dashed) and a
coherent background (magenta dotted) to the
spin-density matrix.
The phase differences between the
exotic wave and the $1^{++}0^+\,\rho\pi\,S$ and the
$2^{-+}0^+\,f_2\pi\,S$ waves have been examined. While the latter
appears to be rather flat, 
%which is
%attributed 
%to the fact that there are two resonances, $\pi_1(1600)$ and $\pi_2(1670)$,
%with very similar masses and widths, causing the relative phase difference 
%to be almost constant. 
%In contrast to this the phase difference to the $1^{++}$
%wave, 
the former, 
shown in 
Fig.~\ref{fig:Pb.3pi.1-+.phase}, 
clearly shows an increase around $1.7\,\GeV/c^2$, suggesting a 
resonant contribution to the $1^{-+}$ wave. 
Although allowed in the model, no significant intensity was found
in the negative-reflectivity spin-exotic wave. 

A much bigger data set was
taken by COMPASS with a liquid hydrogen target, surpassing the
existing world statistics 
by a factor of more than 20. In addition to the $\pi^-\pi^-\pi^+$
final state with 
approximately $100\,\mathrm{M\ events}$ \cite{Haas:2011rj}, 
also the one containing neutral pions,
$\pi^-\pi^0\pi^0$, with more 
than $2.4\,\mathrm{M\ events}$
has been analyzed \cite{Nerling:2012ei}. 
Figure~\ref{fig:H.3pi.1-+.intensity}
shows the intensity in the exotic wave for both final states, where
the intensity in the $2^{++}1^+\,\rho\pi\,D$ wave was used to
normalize the two data sets. A set of 53 waves including a flat wave
was used to account for the increase of statistics.    
Similarly to the Pb target, a peak structure is observed between
$1.6$ and $1.7\,\GeV/c^2$ for the H target. The phase difference to
the $1^{++}$ wave, 
shown in Fig.~\ref{fig:H.3pi.1-+.phase} for the H target, also
exhibits a 
clear rising variation, consistent with a resonant
interpretation of the peak. The structures at lower masses, visible
for the H target, were found to be unstable and largely depending on
the fit model.
%they disappear in an ongoing analysis when thresholds for
%certain isobars, originally introduced because of a lack of
%statistics, are removed, 
%better parameterizations of isobars are used, 
%and the fit is performed in bins of
%4-momentum transfer $t$.   
The intensity in the peak structure at 
$1.67\,\GeV/c^2$ appears to be enhanced with respect to the broad
remaining non-resonant background for the Pb target compared to the H target. 
This observation is consistent with a general enhancement of the
intensity of waves with spin projection $M=1$ and a corresponding
reduction of waves with $M=0$ for the Pb target, observed also for the
dominant $1^{++}$ and $2^{-+}$ waves \cite{Haas:2011rj}. 
For the H target, no fit of the spin
density matrix has yet been performed.   

In order to better understand the background, non-resonant
processes like the Deck effect \cite{Deck:1964hm}
have to be 
taken into  
account in the fit model. As a first step, pure Deck-like events were
generated,  
%according to \cite{Ascoli:1974sp}, 
and the events were
passed 
through the Monte-Carlo simulation framework of the experiment and
were reconstructed and analyzed in the same way as real experimental
data. The 
intensity appearing in the spin-exotic $1^{-+}1^+\,\rho\pi\,P$ wave
was found to reproduce both the 
intensity and the shape of the 
low-mass background in 
Fig.~\ref{fig:3pi.1-+.intensity},   
indicating a large non-resonant contribution to
the observed $1^{-+}1^+\,\rho\pi\,P$ wave. Pure Deck-like events,
however, do not exhibit any enhancement between $1.6$ and
$1.7\,\GeV/c^2$, nor any significant phase motion. A complete
analysis is under way, which includes the Deck amplitude in the fit to
the spin density matrix, and hence allows for interferences between
resonant and non-resonant amplitudes. 

\begin{figure}[tbp]
  \begin{center}
    \subfloat[]{
      \includegraphics[width=0.33\columnwidth]{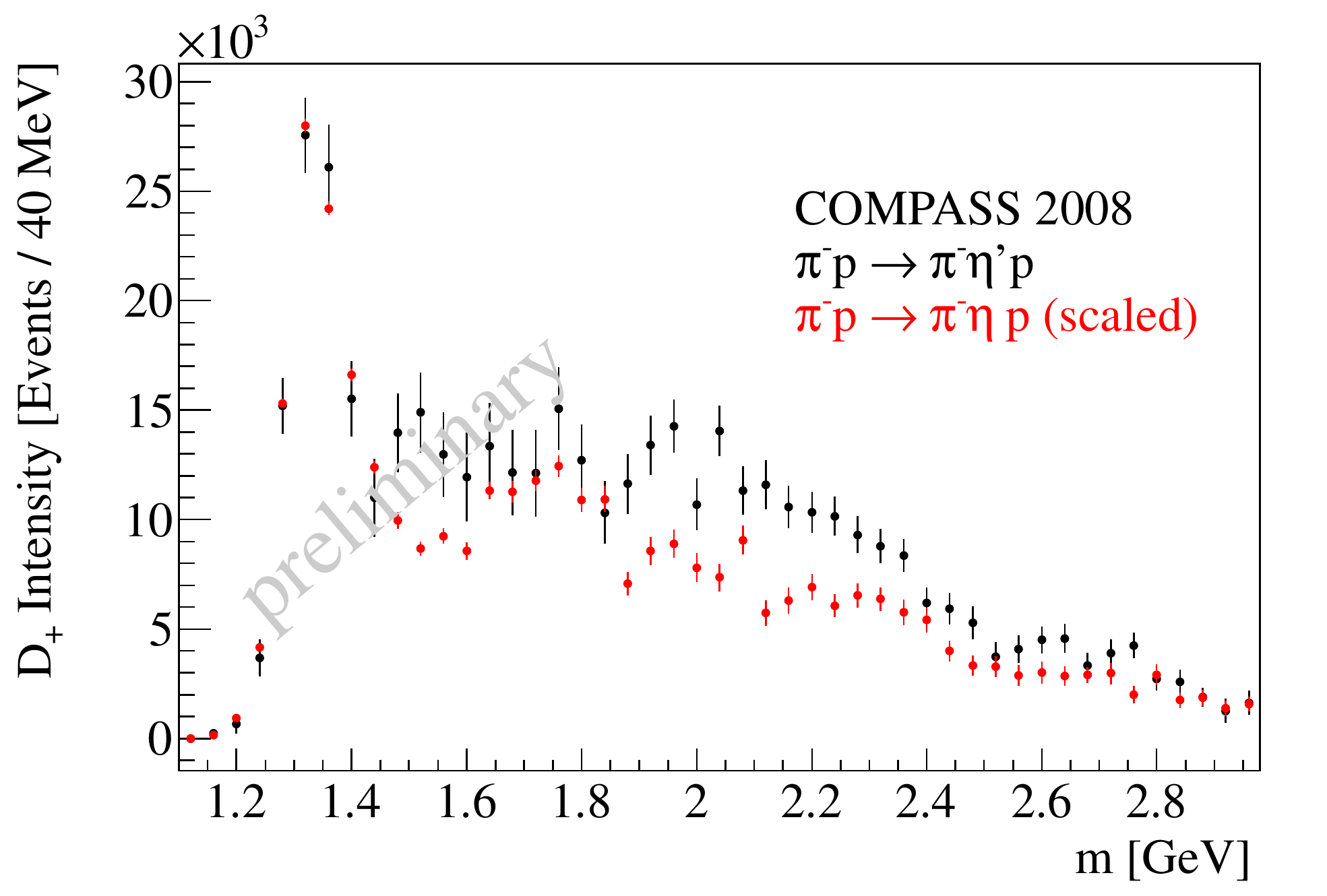}    
      \label{fig:H.etapi.2++.intensity}}
    \subfloat[]{
      \includegraphics[width=0.33\columnwidth]{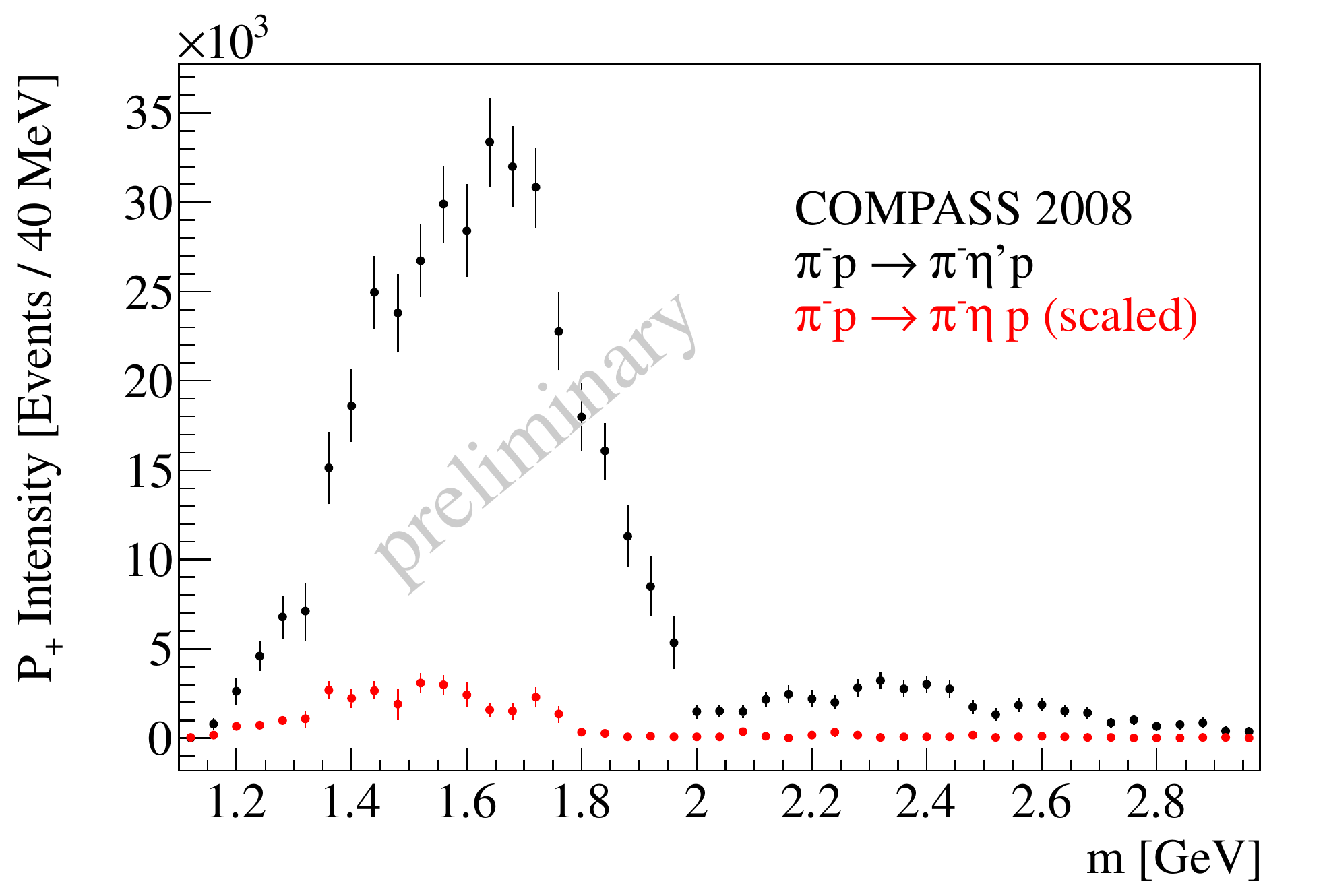}    
      \label{fig:H.etapi.1-+.intensity}}
    \subfloat[]{
      \includegraphics[width=0.33\columnwidth]{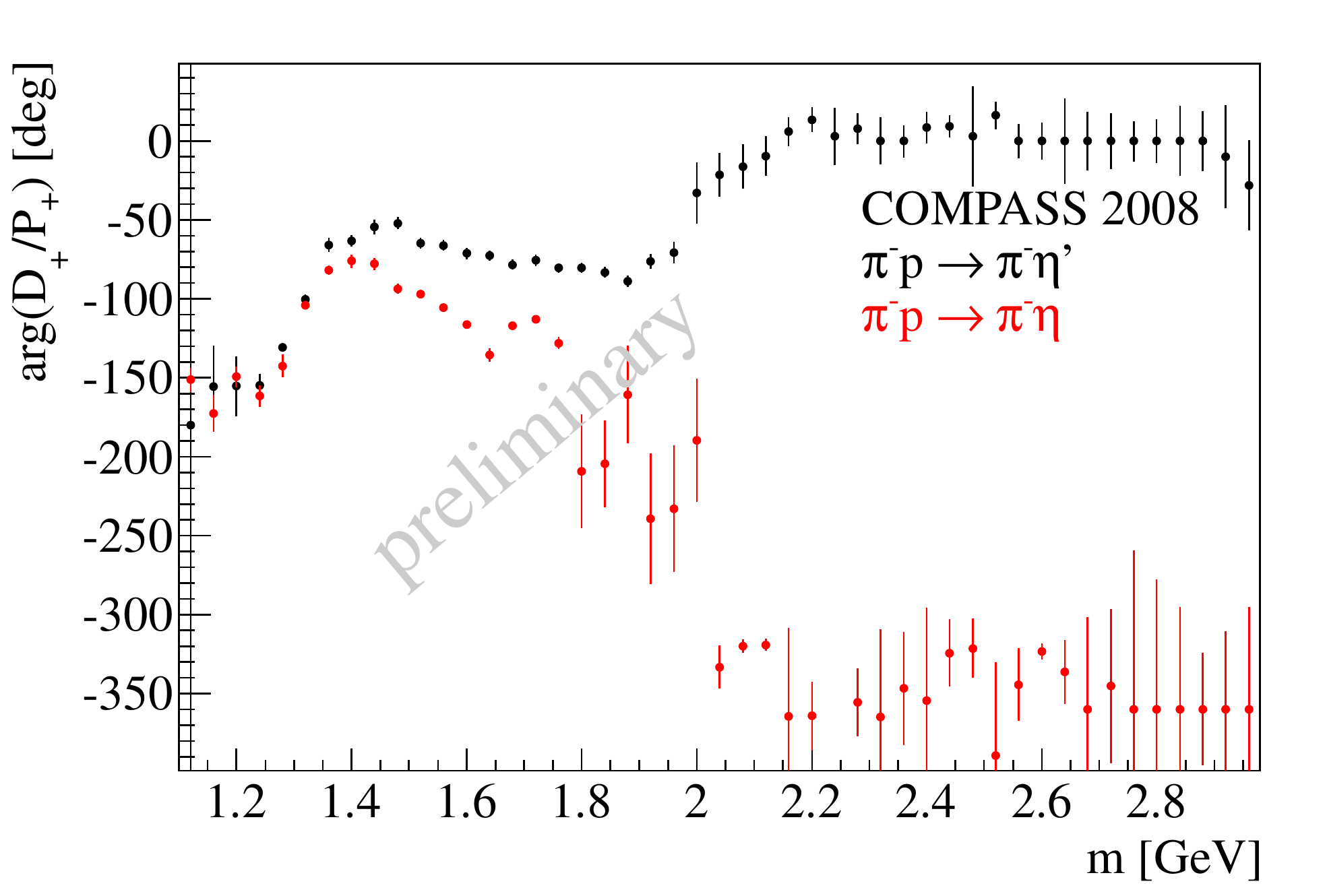}
      \label{fig:H.etapi.1-+.phase}}
    \caption{Comparison of waves for $\eta\pi$ (red data points) and
      $\eta'\pi$ (black data points) final  
      states. (a) Intensity of the $J^{PC}=2^{++}$ $D$ wave, (b)
      intensity of the spin-exotic $1^{-+}$ $P$ wave, (c) phase
      difference between $D$ and $P$ wave.} 
    \label{fig:H.etapi}
  \end{center}
\end{figure}

\paragraph{$\eta\pi$ and $\eta'\pi$ Final States}\mbox{}\\
\indent COMPASS has also analyzed data for $\eta\pi$
($\eta\rightarrow\pi^+\pi^-\pi^0$) and $\eta'\pi$
($\eta'\rightarrow\pi^+\pi^-\eta$, $\eta\rightarrow\gamma\gamma$) 
final states from diffractive scattering of
$190\,\GeV/c$ $\pi^-$ off a H target \cite{Schluter:2012re}. 
In total, about $116\,\mathrm{k\ events}$ were collected for the first
final state, and about $39\,\mathrm{k\ events}$ for the second,
exceeding the statistics of previous 
experiments by more than a factor of $5$. 
%For a 2-scalar final state,
%the quantum numbers are restricted
%by conservation of $P$ and $C$ to 
%$J^{PC}=0^{++}$ ($S$ wave), $1^{-+}$ ($P$ wave, exotic), $2^{++}$ ($D$
%wave),
%$3^{-+}$ ($F$ wave, exotic), 
%$4^{++}$ ($G$ wave) ,$\ldots$. 
Four waves with natural parity exchange, $1^{-+}1^+$ ($P$ wave, exotic),
$2^{++}1^+/2^+$ ($D$ wave), and $4^{++}1^+$ ($G$ wave), were included
in the fit, together 
with a flat background added incoherently. Waves with unnatural parity
exchange were also included, but remained compatible with zero up to
masses of about 
$2.5\,\GeV/c^2$. Figures~\ref{fig:H.etapi.2++.intensity} and 
\ref{fig:H.etapi.1-+.intensity} show the intensities in the
$2^{++}1^+$ and 
$1^{-+}1^+$ waves for the $\eta'\pi$ (black data points) and the
$\eta\pi$ final state (red data points), respectively. Here, the data
points for the latter final state have been scaled by a phase-space
factor $(q'/q)^{J+1/2}$, with $q^{(\prime)}$ being the
breakup momentum into 
$\eta^{(\prime)}\pi$ at a given invariant mass. While the intensities in
the $D$ wave (as well as in the $G$ wave, not shown here) are
remarkably similar in intensity and shape in both final states after
normalization, the $P$ 
wave intensity appears to be very different. For $\eta\pi$, it is
strongly suppressed, while for $\eta'\pi$ it is the dominant wave. The
phase differences between the $2^{++}1^+$ and the $1^{-+}1^+$ waves
are displayed in Fig.~\ref{fig:H.etapi.1-+.phase} for the two final
states. For masses below $1.4\,\GeV/c^2$ the phase differences agree for
the two final states, showing a rising behavior due to the resonating
$D$ wave. For higher masses, the phase differences evolve quite
differently, suggesting a different resonant contribution in the two
final states. 
%At masses above $2.2\,\GeV/c^2$ the phase
%differences seem to agree again (modulo a 180 degree ambiguity in the
%relative phases for the 2-scalar system). 
As for the $3\pi$ final
states, both resonant and non-resonant contributions to the exotic
wave have to be included 
in a fit to the spin-density matrix 
in order to describe both intensities and
phase shifts \cite{Schluter:2012re}. 
Regardless of this, the spin-exotic contribution to the
total intensity is found to be much larger for the $\eta'\pi$ final
state than for the $\eta\pi$ final state, as expected for a hybrid
candidate. 

\subsubsection{Photoproduction}
\label{sec:exp.exotic.photoproduction}
\begin{figure}[tbp]
  \begin{center}
    \subfloat[]{
      \includegraphics[width=0.45\columnwidth]{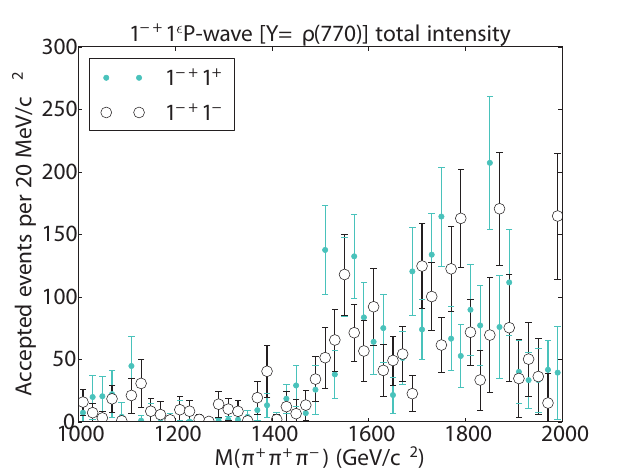}    
      \label{fig:gH.3pi.1-+.intensity}}
    \subfloat[]{
      \includegraphics[width=0.45\columnwidth]{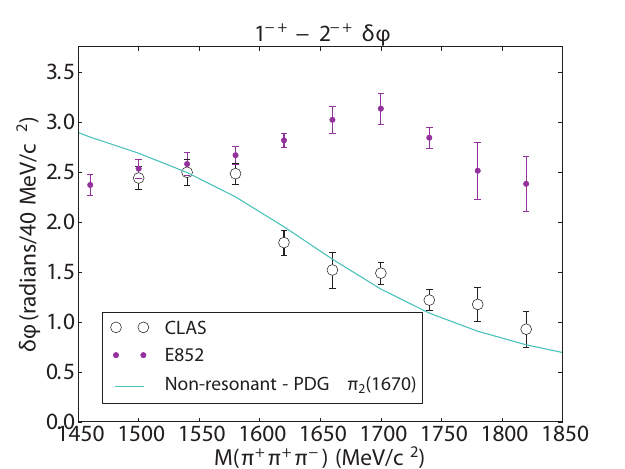}
      \label{fig:gH.3pi.1-+.phase}}
    \caption{The $1^{-+}1^\pm\,\rho\pi\,P$ waves from photoproduction
      at CLAS \cite{Bookwalter:2011cu}. (a) Intensities as a 
      function of $3\pi$ invariant mass, (b) phase difference to the
      $2^{-+}1^\pm\,f_2\pi\,S$ wave, with the corresponding result
      from diffractive production at E852 \cite{Chung:2002pu} overlaid.}
    \label{fig:gH.3pi.1-+}
  \end{center}
\end{figure}

The CEBAF Large Acceptance Spectrometer (CLAS) \cite{Mecking:2003zu}
at Hall B of Jefferson 
Laboratory is studying photo- and electro-induced hadronic reactions
by detecting final states containing charged and neutral
particles. Two experimental campaigns in 2001 and 2008 
%g6c in 2001 and g12 in 2008
were dedicated to the search for exotic mesons photoproduced in the
charge exchange reaction $\gamma p\rightarrow\pi^+\pi^+\pi^- (n)$ (see
Fig.~\ref{fig:exp.production.photo}). A 
tagged-photon beam produced from an electron beam with energies up to 
$5.75\,\GeV$ was impinging on an $18\,\Cm$ long liquid hydrogen
target. 
In order to increase the acceptance for meson production, the target
was shifted upstream by $1\,\m$ and the toroidal magnetic field was
reduced to half its nominal value. Charged particles were detected by a
toroidal spectrometer equipped with drift 
chambers and identified by a time-of-flight detector.
Events with three 
outgoing pions were selected by vertex and timing cuts. The recoiling
neutron was identified through its missing mass. 
The first campaign resulted in a set of $83\,\mathrm{k\ events}$
subjected to a partial wave analysis \cite{Nozar:2008bea}. No evidence
for a $1^{-+}$ 
exotic state, nor for the $1^{++}$ state was found. The second data
taking period yielded a much larger data set of $520\,\mathrm{k\
  events}$ for the PWA \cite{Bookwalter:2011cu}, after the background
from  
baryon resonances in 
the target vertex had been removed by cuts on 4-momentum transfer 
$t'<0.105\,\GeV^2/c^2$  and
on the laboratory angle of both $\pi^+$,
$\theta_\mathrm{lab}(\pi^+)\leq 25^\circ$, to select peripheral events
with forward-going pions. 
%These cuts were shown by MC simulations to remove most of the baryonic
%background without significantly affecting the mesonic sample. 
The PWA included 19 waves with $J^{PC}=1^{++}$, $2^{++}$, $1^{-+}$,
and $2^{-+}$, and a flat background wave. Helicity conservation forbids
the production of waves 
with $M=0$, and hence $J=0$, for an incoming photon and pion
exchange. Waves with $M^\epsilon=1^\pm$ are expected to be populated
equally as a result of ambiguities.  
Evidence was found for the $a_1(1260)$, the $a_2(1320)$ and the
$\pi_2(1670)$ in the 
$1^{++}1^{\pm}\,\rho\pi\,S$ wave, the $2^{++}1^{\pm}\,\rho\pi\,D$ wave and the
$2^{-+}1^{\pm}\,f_2\pi\,S$ wave,
respectively. A non-negligible population of the $M=0$ spin projection of
the $1^{++}$ and $2^{-+}$ waves, however, indicates a possible leakage or
non-resonant $S$ wave production via the Deck effect. 
The intensity of the exotic $1^{-+}1^\pm\,\rho\pi\,P$ wave, shown in
Fig.~\ref{fig:gH.3pi.1-+.intensity} as a function of the $3\pi$
invariant mass, does not exhibit any evidence for structures
around $1.7\,\GeV/c^2$. The fluctuations at masses above
$1.8\,\GeV/c^2$ are not understood yet. 
Its phase difference relative to the
$2^{-+}1^\pm\,f_2\pi\,S$ wave, displayed in
Fig.~\ref{fig:gH.3pi.1-+.phase}, shows a continuous downward trend in the
relevant mass region, as expected for a $2^{-+}$ resonance subtracted
from a non-resonant background (blue line). 
% Also shown in
% Fig.~\ref{fig:gH.3pi.1-+.phase} is the phase difference observed in
% \cite{Chung:2002pu}, which shows a slight upward trend. As noted in
% Sec.~\ref{sec:exp.exotic.diffraction}, the COMPASS experiment does not see
% any significant phase motion of the $1^{-+}$ wave with respect to the
% $2^{-+}$ wave, and 
% attributed this to the presence of two resonances with very similar
% mass and width. 
The conclusion from the CLAS experiments is that there is no evidence
for an exotic $1^{-+}$ wave in photoproduction,  
with an upper limit of 
$2\%$ of the total intensity. 

\begin{figure}[tbp]
  \begin{center}
    \subfloat[]{
      \includegraphics[width=0.45\columnwidth]{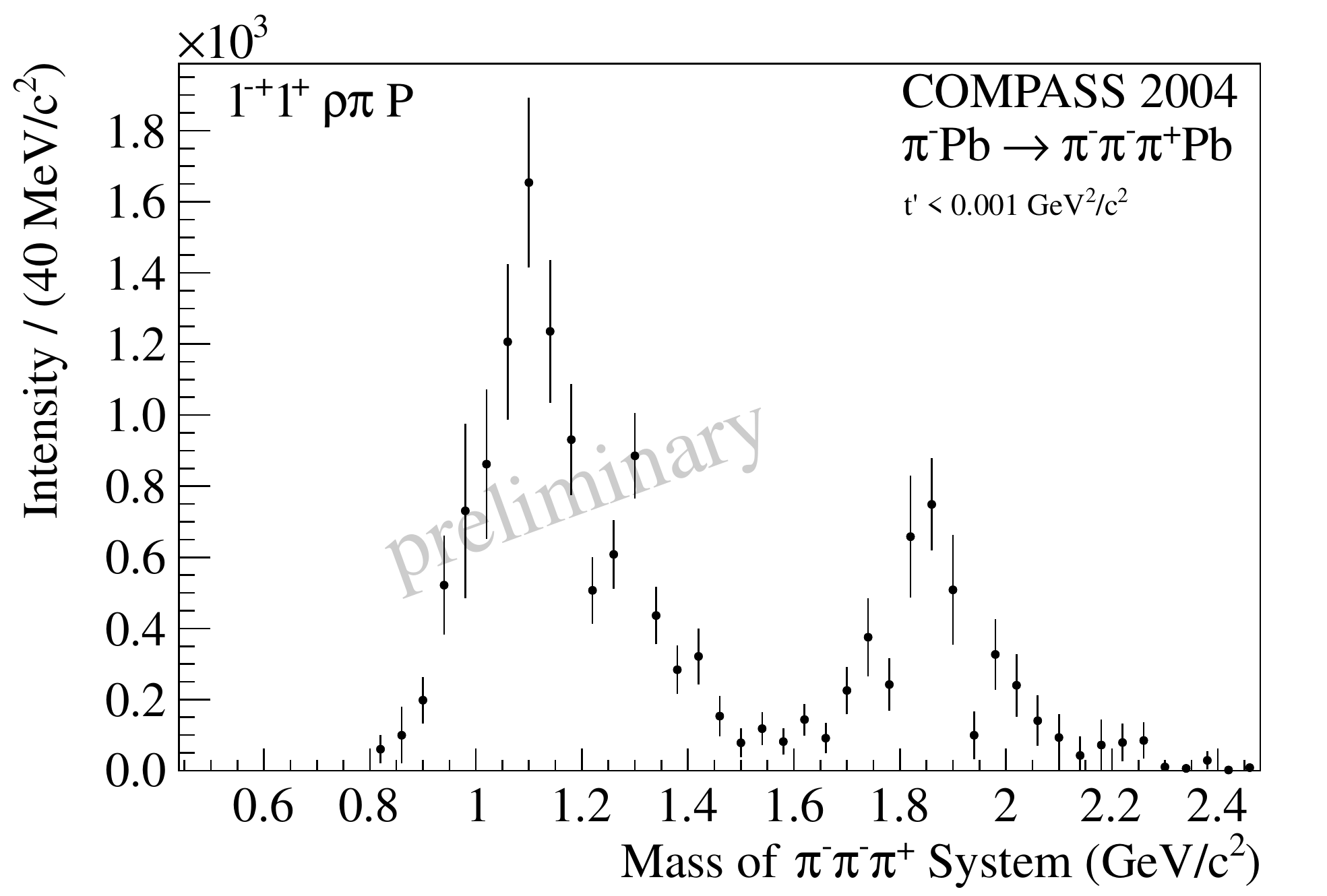}    
      \label{fig:Pb.3pi.lowt.1-+.intensity}}
    \subfloat[]{
      \includegraphics[width=0.45\columnwidth]{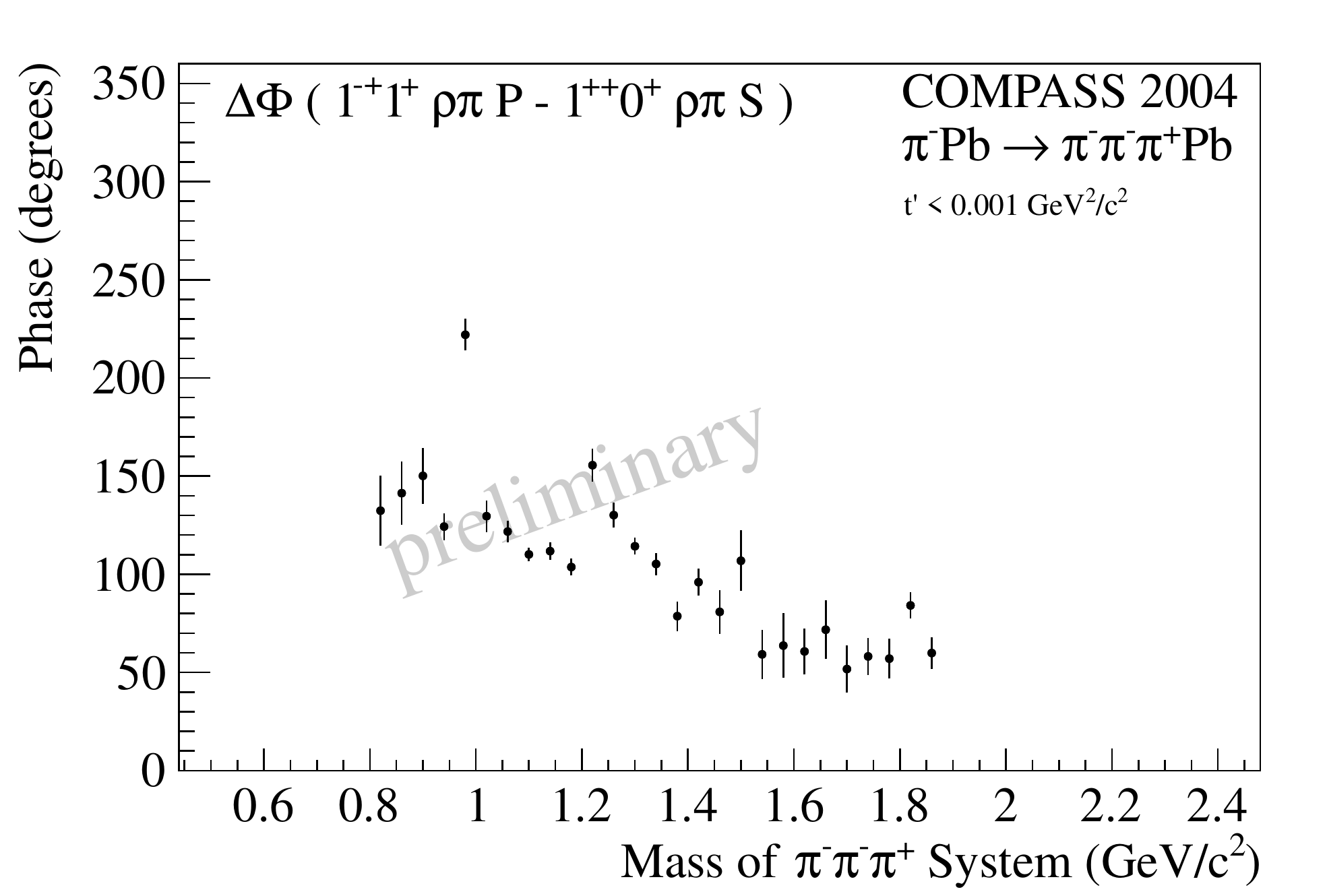}
      \label{fig:Pb.3pi.lowt.1-+.phase}}
    \caption{The $1^{-+}1^\pm\,\rho\pi\,P$ waves from photoproduction
      at COMPASS. (a) Intensity
      vs. $3\pi$ invariant mass, (b) phase relative to the
      $1^{++}0^+\,\rho\pi\,S$ wave.}
    \label{fig:Pb.3pi.lowt.1-+}
  \end{center}
\end{figure}

The COMPASS experiment studied pion-induced reactions on a Pb target
at very low values of 4-momentum transfer, $t'<0.001\,\GeV^2/c^2$,
which can proceed both diffractively via the exchange of a Pomeron and via 
the exchange of quasi-real photons from the Coulomb field of the
heavy nucleus (see
Fig.~\ref{fig:exp.production.coulomb}). 
A partial wave analysis of
this data set consisting of approximately $1\,\mathrm{M\ events}$ was
performed using a total of 37 waves, 
showing no sign of a resonance in the $1^{-+}1^\pm\,\rho\pi\,P$ wave
at a mass of 
$1.7\,\GeV/c^2$ (c.f. Fig.~\ref{fig:Pb.3pi.lowt.1-+}), consistent with
the CLAS observation.  

\subsection{Non-exotic Hybrids}
\label{sec:exp.nonexotic}
While there is some evidence for an isovector member of a light
$1^{-+}$ exotic 
nonet, as detailed in Sec.~\ref{sec:exp.exotic.diffraction}, members
of non-exotic multiplets will be more difficult to identify. Most of
the light meson resonances observed until now are in fact compatible
with a $q\overline{q}'$ interpretation. 
Taking the LQCD predictions as guidance, the lowest
isovector hybrids with ordinary quantum numbers should have
$J^{PC}=0^{-+}$, $1^{--}$, and $2^{-+}$ (see
Sec.~\ref{sec:theory.lqcd}). We will review the latest observations
concerning resonances with these quantum numbers. 
%The final test for
%the hybrid hypothesis of these candidate states of course will be the
%identification of the isoscalar and strange members of a multiplet. 

\begin{figure}[tbp]
  \begin{center}
    \subfloat[]{
      \includegraphics[width=0.5\columnwidth]{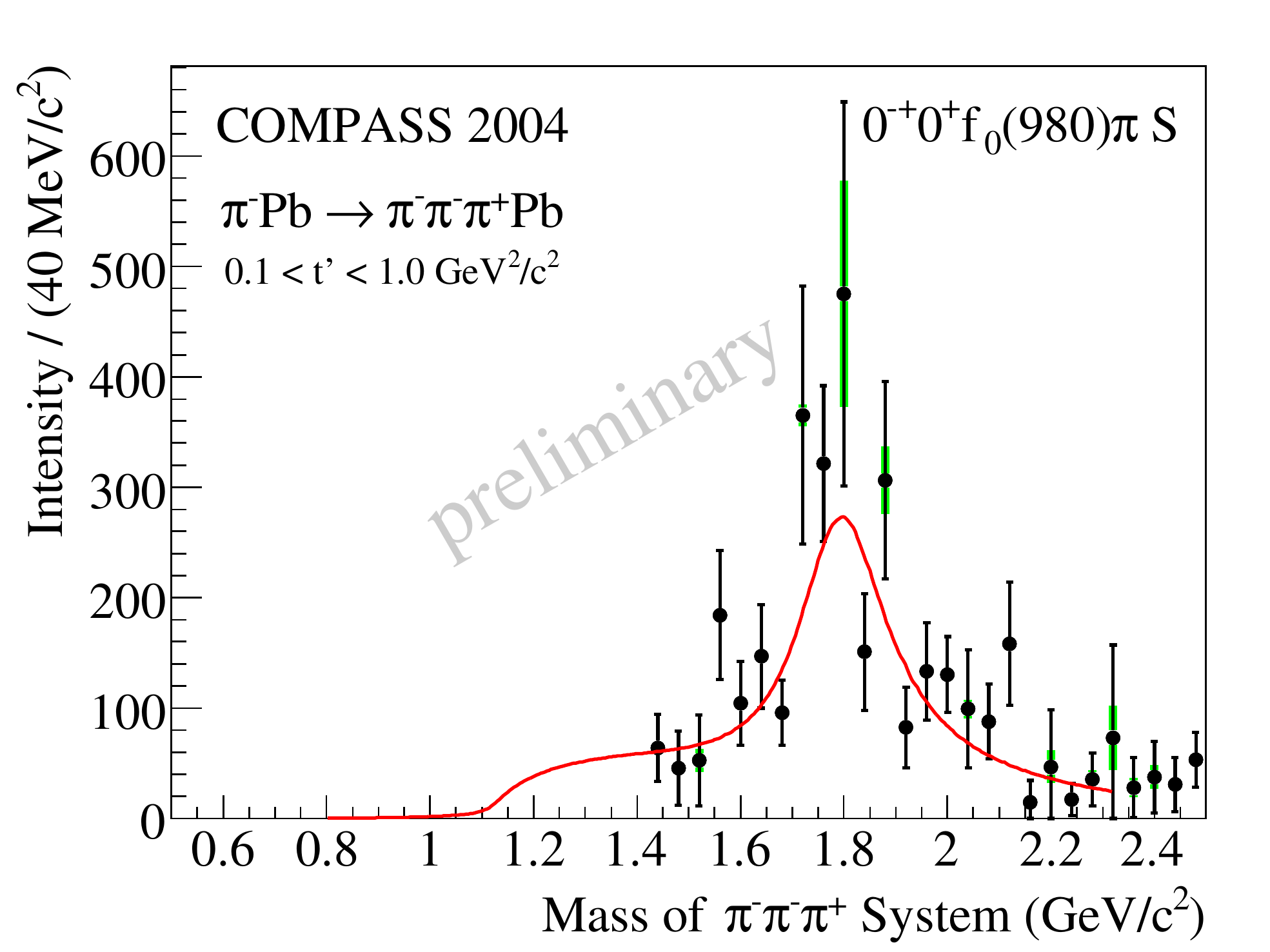}    
      \label{fig:Pb.3pi.0-+.intensity}}
    \subfloat[]{
      \includegraphics[width=0.4\columnwidth]{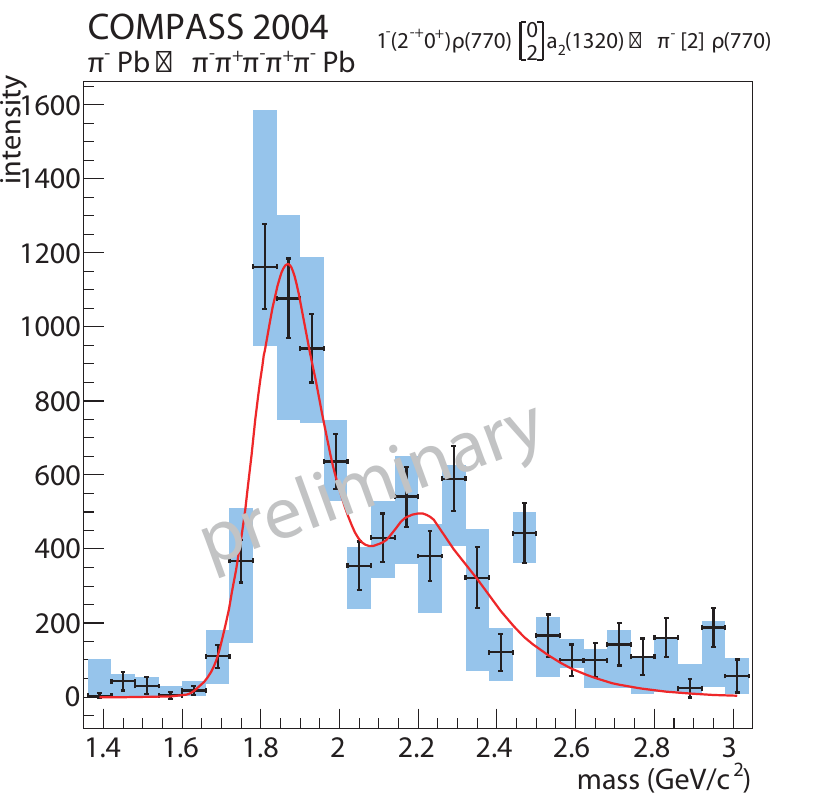}    
      \label{fig:Pb.5pi.2-+.intensity}}
    \caption{Intensities of (a) the $0^{-+}0^+\,f_0(980)\pi\,S$ wave
      in the $3\pi$ final state, and (b) 
      the
      $2^{-+}0^+\,\rho a_2(1320)[\rho\pi]\,D$ wave in the $5\pi$ final
      state, both from a Pb target.}
    \label{fig:Pb.0-+.2-+.intensity}
  \end{center}
\end{figure}

\paragraph{The $\pi(1800)$}\mbox{}\\
There is clear experimental evidence for the $\pi(1800)$ 
%The PDG
%gives 
%$M=(1812\pm 12)\,\MeV/c^2$ and 
%$\Gamma=(208\pm 12)\,\MeV/c^2$
%for the mass and width of the $\pi(1800)$, respectively
\cite{Beringer:2012zz}. The latest measurements of this state come from the
COMPASS experiment, using a $190\,\GeV/c$ $\pi^-$ beam impinging on a
Pb target. Figure~\ref{fig:Pb.3pi.0-+.intensity} shows the intensity in
the dominant $0^{-+}$ wave 
determined from a partial wave analysis of the
$3\pi$ final state. 
A strong $\pi(1800)$ signal was recently also observed 
in an analysis of the $5\pi$  
final state from COMPASS \cite{Neubert:2010zz}. 
Table~\ref{tab:Pb.0-+.2-+.results} summarizes
the masses and 
widths obtained by fitting Breit-Wigner functions to
the spin density matrix. 
\begin{table}[tbp]
  \caption{Mass and width of the $\pi(1800)$ and $\pi_2(1880)$, 
    obtained from recent 
    partial wave analyses of the COMPASS experiment.} 
  \label{tab:Pb.0-+.2-+.results}
  \centering{\small
  \begin{tabular}{ccccc} \hline\hline
    Resonance & $J^{PC}M^\epsilon\,[\mathrm{channel}]\,L$ & Mass
    ($\MeV/c^2$) & Width ($\MeV/c^2$) & Events \\ \hline
    $\pi(1800)$ & $0^{-+}0^+\,f_0(980)\pi\,S$ & $1785\pm 9^{+12}_{-6}$ 
    & $208\pm 22^{+21}_{-37}$ & $420\,\mathrm{k}$\\ 
    & $0^{-+}0^+\,f_0(1500)[\rho\rho]\pi\,S$ & $1781\pm 5^{+1}_{-6}$ 
    & $168\pm 9^{+5}_{-14}$ & $200\,\mathrm{k}$ \\ \hline
    $\pi_2(1880)$ & $2^{-+}0^+\,\rho a_2(1320)[\rho\pi]\,D$ & $1854\pm
    6^{+6}_{-4}$  
    & $259\pm 13^{+13}_{-17}$ & $200\,\mathrm{k}$ \\ \hline\hline 
  \end{tabular}
}
\end{table}
The interpretation of the experimental decay pattern is not
straightforward.  
On the one hand, it prefers to decay into a pair of $L=1$ and $L=0$
mesons, as expected 
for a hybrid (c.f. Sec.~\ref{sec:theory.decay} and
Table~\ref{tab:hybrid.decay})). Large partial 
widths for 
$f_0\pi$ decays, as expected for a hybrid state, have been observed  
by COMPASS and others \cite{Beringer:2012zz}.  
The 
decay of a $3S$ $q\overline{q}'$ state into this final state should be
small. 
On the 
other hand, both the hybrid and $q\overline{q}'$ state are predicted to
have a sizable $\rho\pi$ partial width, which is not seen.  
%The decay into $\omega\rho$ should be 
%suppressed for a hybrid, but has been observed by one experiment
%\cite{Amelin:1999gk} so far, albeit at a lower mass. This channel is
%not included in the PDG averages.   
%Therefore, the
%suggestion 
%has been made that two distinct states have been observed
%\cite{Barnes}, an opinion which is not shared by all
%\cite{Klempt}. 
More statistics and advanced coupled-channel analyses
are certainly needed to clarify the situation.  

\paragraph{The $\pi_2(1880)$}\mbox{}\\
There is growing experimental evidence for the existence of this
state. 
%The PDG lists as parameters of the 
%$\pi_2(1880)$ a mass of  
%$M=(1895\pm 16)\,\MeV/c^2$ and a width of $\Gamma=(235\pm
%34)\,\MeV/c^2$.
The latest high-statistics measurements of this state again
come from COMPASS. 
A clear
peak is observed in the 
intensity of the $2^{-+}0^+\,f_2\pi\,D$ wave for both a Pb and a H
target \cite{Haas:2011rj}, which is 
shifted in mass with respect to the $\pi_2(1670)$, and also exhibits a
phase motion relative to the latter in the $f_2\pi\,S$ wave. This
observation, however, was also explained 
differently, including e.g.\  
the interference of the $f_2\pi\,S$ wave with a Deck-like amplitude,
which shifts the true $\pi_2$ peak to lower masses \cite{Dudek:2006ud}.  
For $5\pi$ final states \cite{Neubert:2010zz}, a total of three
resonances are needed to 
describe the $2^{-+}$ sector, the $\pi_2(1670)$, the $\pi_2(1880)$,
and a high-mass $\pi_2(2200)$. Figure~\ref{fig:Pb.5pi.2-+.intensity}
shows the intensity in the $2^{-+}0^+\,\rho a_2(1320)[\rho\pi]\,D$
wave, together with a fit of three Breit-Wigner functions to the
spin-density matrix. 
The resulting mass and width deduced from this fit are 
given in Table~\ref{tab:Pb.0-+.2-+.results}.
%A possible isoscalar partner of the $\pi_2(1880)$, the $\eta_2(1870)$
%has also been reported \cite{Beringer:2012zz}, but needs confirmation. 

\paragraph{States with $J^{PC}=1^{--}$}\mbox{}\\
The PDG lists two $\rho$-like excited states, the
$\rho(1450)$ and the $\rho(1700)$, observed in $e^+e^-$ annihilation,
photoproduction, antiproton annihilation and $\tau$ decays
\cite{Beringer:2012zz}. Their masses are  
consistent with the $2^3S_1$ and $1^3D_1$ $q\overline{q}'$ states,
respectively, but their decay
patterns do not follow the $^3P_0$ rule \cite{Barnes:1996ff}. The
existence of a light vector hybrid state, mixing with the
$q\overline{q}'$ states, was proposed to solve
these discrepancies \cite{Donnachie:1999re}. 
Recently, BaBar has reported the observation of a $1^{--}$ state
decaying to $\phi\pi^0$ \cite{Aubert:2007ym}, the $\rho(1570)$. 
%, which might be identical
%to an earlier observation in Serpukhov \cite{Bityukov:1986yd}. 
Interpretations of this signal include a new state, a
threshold effect, and an OZI-suppressed decay of the $\rho(1700)$. 
%A very broad vector state with pole position
%$M=(1576^{+49+98}_{-55-91}+\frac{i}{2}
%818^{+22+64}_{-23-133})\,\MeV/c^2$ has been reported by BES
%\cite{Ablikim:2006hp}. It has been interpreted to be due to interference
%effects in final state interactions, and in tetraquark scenarios. 
In conclusion, there is no clear evidence for a hybrid state with
vector quantum numbers. A clarification of the nature of the
$\rho$-like states, especially above $1.6\,\GeV/c^2$, requires
more data. 

\subsection{Heavier Candidates}
\label{sec:exp.heavy}
To avoid experimental difficulties in the light quark sector due to the high
density of ordinary $q\overline{q}'$
states 
below $2.5\,\GeV/c^2$, a search for hybrids in the less populated
charmonium mass region is expected to be very rewarding. Model calculations as 
well as LQCD predict the ground state of the charmonium hybrid at a
mass of around $4.3\,\GeV/c^2$ with exotic quantum numbers $1^{-+}$
\cite{Isgur:1985vy,Chen:2000qj,Juge:2003qd}.
%, excluding mixing effects with nearby \Pqc\Paqc\ states. 
%A dynamical
%selection rule forbidding the decay into two open charm mesons with
%$L=0$ is part of various models, and suggests that this state could be
%narrow with a width between $5$ and $50\,\MeV$. 
No candidate state with such quantum numbers, however, has been
identified experimentally until now, although many new and yet
unexplained states containing charm 
and bottom quarks have been discovered (for a recent review, see e.g.\
\cite{Brambilla:2010cs}).  
Anomalous decay properties and an apparent overpopulation with 
respect to expectations for 
$c\overline{c}$ states suggested a non-$q\overline{q}$ nature of 
$Y(4260)$ 
state, a $1^{--}$ state discovered by BaBar in initial 
state radiation (ISR) \cite{Aubert:2005rm}.
Possible interpretations range from a tetraquark
system to a 
charmonium hybrid.  
The $Y(2175)$, also discovered by BaBar in ISR \cite{Aubert:2006bu},
has a decay pattern which is strikingly similar to the $Y(4260)$ and
the $\Upsilon(10860)$,
%\cite{Wang:2012wa}, 
and was therefore discussed as the strangeonium hybrid partner of the
$Y(4260)$ \cite{Ding:2007pc,Zhu:2007wz}. There is, however, no  
overpopulation of states in the $s\overline{s}$ sector, so radial and orbital
excitations have to be considered as well. 

%\subsection{New Experiments}
%\label{sec:exp.new}

\section{Conclusions and Outlook}
\label{sec:conclusion}
The self-interacting nature of the gluons in QCD allows excitations of
the gluonic field, which confines quarks inside mesons, to contribute
to the quantum numbers of hadrons. 
Models of QCD predict the
lowest-lying 
hybrid states at masses varying between $1.3$ and $2.2\,\GeV/c^2$, and
with different 
quantum numbers and excitation patterns. Recently, a 
fully unquenched 
LQCD calculation yielded predictions for the complete
spectrum of isovector mesons made of light quarks, including a light
hybrid supermultiplet with quantum numbers $J^{PC}=0^{-+}$, $1^{--}$,
$2^{-+}$, and a spin-exotic $1^{-+}$. 

The experimental observation of hybrid mesons thus constitutes a
stringent test of non-per\-tur\-ba\-tive QCD. 
They can be identified either by an
overpopulation of states in a given mass region with respect to
quark model expectations, or by the detection of states with exotic
quantum numbers. 
A final proof of their nature of course can only come from the
observation of  
the decay pattern and from branching ratios, which are expected to be
different for hybrids from the one of ordinary mesons, and from the
identification of isoscalar, isovector, and strange members of a
multiplet.  
In addition, the production 
mechanism is an important question which can be clarified only by
studying the systems using different probes and at different
energies, employing polarization variables where possible.  

The COMPASS experiment
has provided clear evidence for the presence of a spin-exotic wave 
with quantum numbers $J^{PC}=1^{-+}$ in $\rho\pi$ and $\eta'\pi$ final
states in diffractive production from a $190\,\GeV/c$ 
pion beam, consistent with the $\pi_1(1600)$, although non-resonant
production seems to be playing an important role. 
CLAS as well as COMPASS have shown that photoproduction of a $1^{-+}$
hybrid does not occur with the expected strength. By
studying the 
reaction $\gamma p\rightarrow\pi^-\pi^+\pi^+ (n)$, CLAS has set an upper limit 
for the $1^{-+}$ wave of $2\%$ of the total intensity, suggesting that
the production of the $\pi_1(1600)$ proceeds via Pomeron exchange,
but is suppressed in charged-pion exchange photoproduction. This
conjecture  
calls for a better understanding of the production mechanism, e.g.\ by
analyzing 
its dependence on 4-momentum transfer or center-of-mass
energy. 

In addition to the spin-exotic $\pi_1(1600)$, there are 
candidates for supernumerary states with ordinary quantum numbers
$0^{-+}$ and 
$2^{-+}$, the $\pi(1800)$ and the $\pi_2(1880)$, respectively. 
Both states have been 
observed in the past, with new evidence coming from COMPASS in $3\pi$
and $5\pi$ final states. 
These states would account for the full
set of light hybrid states with the $q\overline{q}'$ pair in $S=1$ 
configuration, as predicted from LQCD and quasiparticle models. 
Candidates for additional states in the $\rho$-like sector, which
could make up the spin-singlet member of the supermultiplet, 
have been observed at $e^+e^-$ machines, but require confirmation.   

New hadron spectroscopy experiments are on the horizon,
which are expected to considerably advance our 
understanding of the meson spectrum. Key features
of these experiments 
will be high statistics requiring highest possible luminosities and
sensitivity to production cross sections in the sub-nanobarn
region. This can only be achieved by hermetic detectors with excellent
resolution and particle 
identification capabilities, providing a very high acceptance for charged
and neutral particles. 
BESIII at the BEPC-II $e^+ e^-$ 
collider in Beijing has already started to take data in the
$\tau$-charm region at a luminosity of $\EE*{33}\,\Cm^{-2}\,\s^{-1}$,
with a maximum center-of-mass energy of $4.6\,\GeV$.   
At Super-B factories, aiming at a 100-fold luminosity increase to values
of 
$\sim\EE*{36}\,\Cm^{-2}\,\s^{-1}$, 
the sensitivity for new states in the charm and bottom sector will 
increase dramatically.  
%Experiments at the LHC, especially LHCb with its excellent resolution,
%are also expected to deliver 
%high-statistics data on the meson spectrum. 
GlueX is a new experiment which will study photoproduction of mesons
with masses below $3\,\GeV/c^2$ at the $12\,\GeV$ upgrade of CEBAF,
Jefferson Laboratory. An important advantage of the experiment will be
the use of polarized photons, which narrows down the possible initial
states and gives direct information on the production process. 
PANDA, a new experiment at the FAIR antiproton storage ring HESR, is
designed for high-precision studies of the 
hadron spectrum in the charmonium mass range. 
% In $\overline{p}p$
% annihilations, all states with 
% non-exotic quantum numbers can be formed directly. Consequently, the
% mass resolution 
% for these states is only limited by the beam momentum
% resolution.
% %Advanced cooling techniques (stochastic and electron
% %cooling) provide \Pap\ beams with a relative momentum resolution
% %of $\EE*{-5}$, which directly translates into mass resolution when very
% %thin targets 
% %% like a hydrogen gas jet or hydrogen pellets 
% %are used.  
% Spin-exotic states can be obtained in production experiments. 
It is expected to run at center-of-mass energies between $2.3$ and
$5.5\,\GeV$ 
with a maximum luminosity of $2\EE{32}\,\Cm^{-2}\,\s^{-1}$. 
%Therefore,
%a signature providing clear evidence for such a state is that it is
%observed in production, but not in formation. 
%In
%addition, the gluon-rich environment of \Pap\Pp\ annihilations is
%considered preferable 
%for the production of hybrids and glueballs. 

\end{document}